\documentclass[]{sig-alternate-10pt}
\usepackage{subcaption}
\usepackage{amsmath}
\usepackage{graphicx}
\usepackage{enumerate}
\usepackage{xspace}
\usepackage{url}
\usepackage{amsmath}
\usepackage[ruled]{algorithm}
\usepackage[noend]{algorithmic}
\usepackage{enumitem}

\usepackage{cite,cases,color,array,url}
\usepackage{algorithmic}
\usepackage{graphicx}
\usepackage{mdwmath}
\usepackage{mdwtab}
\usepackage{eqparbox}
\usepackage{amsopn}
\usepackage{multirow}
\usepackage{graphics}
\usepackage[font=footnotesize]{subfig}
\usepackage{url}
\usepackage{paralist}
\usepackage{xspace}
\usepackage{amsmath}
\usepackage{epsfig}
\usepackage{latexsym}
\usepackage{amsfonts}
\usepackage{amssymb}
\usepackage{comment}
\usepackage{mathrsfs}
\usepackage{cases}

\newcommand{\MYCUT}[1]{{ }}

\def\ie{\textit{i.e.}\xspace}

\def\eg{\textit{e.g.}\xspace}

\newcommand{\eqqref}[1]{Eq.~(\ref{#1})}

\def \ourprotocol{Swadloon\xspace}

\allowdisplaybreaks

\begin{document}

\title{Accurate Indoor Localization Using Acoustic Direction Finding via Smart Phones}
\author{ Wenchao Huang, Yan Xiong 
\thanks{
    Wenchao Huang and Yan Xiong are with the School of Computer Science and Technology, University of Science and Technology of China, email:\{huangwc, yxiong\}@ustc.edu.cn. 
}
, Xiang-Yang Li
\thanks{
    Xiang-Yang Li is with Department of Computer Science, Illinois Institute of Technology, email: xli@cs.iit.edu.
}
, Hao Lin
\thanks{
    Hao Lin is with School of Internet of Things Engineering, Jiangnan University, email: imlinhao@gmail.com.
}
, Xufei Mao
, Panlong Yang
, Yunhao Liu
\thanks{
    Panlong Yang is with Institute of Communication Engineering, PLAUST.
    Xufei Mao and Yunhao Liu are with Department of Software Engineering, Tsinghua University.
}
}
\maketitle

\begin{abstract}
We propose and implement a novel indoor localization scheme,
 \textbf \ourprotocol, built upon an accurate acoustic direction finding.
 \ourprotocol leverages sensors of the smartphone
 without the requirement of any specialized devices.
The scheme \ourprotocol does not rely on any fingerprints and is very
 easy to use: a user only needs to shake the phone for a short
 duration before walking and localization.
Our  \ourprotocol design exploits  a key observation:
 the relative displacement and velocity of the phone-shaking
 movement corresponds to the subtle phase and frequency shift of the
 Doppler  effects experienced in the received acoustic signal by the
 phone.
A novel method is designed to derive the direction from the phone to
 the acoustic source  by combining the velocity calculated from the
 subtle Doppler shift with the one  from the inertial sensors of the
 phone.
Then a real-time precise localization and tracking is enabled by
 using a few anchor speakers with known locations.
Major challenges in implementing \ourprotocol are to measure the
 frequency shift precisely and to estimate the shaking velocity
 accurately when the speed of phone-shaking is low and changes
 arbitrarily.
We propose rigorous methods to address these challenges, and
 then design and deploy \ourprotocol in several  floors of an indoor
 building each with area  about $2000m^2$.
Our extensive experiments show that the  mean
 error of direction finding is around $2.1^o$ when the
 acoustic source is within the range of $32m$.
For indoor localization, the $90$-percentile errors are under $0.92m$,
 while the maximum error is $1.73m$ and the mean is about $0.5m$.
For real-time tracking, the errors are  within $0.4m$ for
 walks of  $51m$.
\end{abstract}
\section{Introduction}
\label{sec:intro}

Phone-to-phone direction finding is attractive in mobile social
 networks nowadays for supporting various applications, \eg,
 friending, and sharing.
Recent mobile apps have made similar functions, such as
 Facebook's Friendshake \cite{friendshake} and Google Latitude
 \cite{GoogleLatitude}. However, they are based on GPS and cannot be applied
 to indoor environment.  An accurate method of direction finding is by
 using directional antenna
\cite{4711074,4509717,Niculescu:2004:VBS:1023720.1023727},
but it requires specialized hardware and clearly limits the
availability to regular users. Several approaches of direction
finding by smartphones have been proposed
\cite{2011-MOBICOM-Iamantenna,2012-MobiQuitous2011-ProposalDirectionEstimation,2011-SenSys-feasibilityrealtime}. However,
it  remains a challenge to do accurate direction finding by phone
 under long distance.

Precise indoor localization is also important for location based
 services.
Those methods achieving high accuracy usually require special hardware
 not readily available on smartphones
\cite{6071927}, or infrastructures
 expensive to deploy
 \cite{DBLP:journals/tsmc/LiuDBL07}.
Pure WiFi-based localization can achieve reasonable accuracy (\eg,
 3$\sim$$4m$), but there always exist large errors (\eg, 6$\sim$8m)
 unacceptable for many scenarios  \cite{2012-MOBICOM-PushlimitWiFi}.
Though there have  been many proposals  improving the accuracy of WiFi
 based localization (\eg, with $80$-percentile errors about $1m$
 \cite{2012-MOBICOM-PushlimitWiFi}) by exploiting additional signals,
 low-cost precise indoor localization is still  challenging.

We propose \textbf \ourprotocol, a
\emph{\textbf{S}hake-and-\textbf{W}alk} \textbf{A}coustic
 \textbf{D}irection-finding and indoor \textbf{LO}calizati\textbf{ON}
 scheme using smartphones.
\ourprotocol has two key components, precise phone-to-phone
 (or phone-to-speaker) direction finding and
 accurate indoor localization, each of which has a wide range of
 applications.
Assume that there is an acoustic signal emitted from a speaker or a
 phone.
\ourprotocol exploits the fact that  shaking the smartphone or walking with
 the smartphone  will cause Doppler effects on the
  acoustic signal received by the smartphone.
\ourprotocol  precisely measures
 the real-time \textit{phase and frequency shift} of the Doppler effect, which
 corresponds to the \textit{relative displacement and velocity} from the phone to the
 acoustic source respectively.
\ourprotocol then obtains the accurate \textit{direction} of the
 acoustic source by combining the relative velocity calculated from the
 Doppler shift with the one from the  inertial sensors of the
 smartphone, \ie, the accelerometer and the gyroscope.

The main challenges of implementing \ourprotocol are the
 noisy data collected from inertial sensors, and the
 measurement of the subtle frequency shift when the motion velocity of
 phone is slow or fluctuates continuously.
We propose several rigorous methods (discussed in detail in
 Section~\ref{sec:directionfinding}) in \ourprotocol to address these
 challenges, \eg, we use Phase Locked Loop (PLL) to
 precisely measure the phase and frequency shift.
Note that for phone-to-phone direction finding,
 the object phone of direction finding serves as an acoustic source,
 and the finder shakes his/her phone gently to produce the Doppler effect.

Based on this precise  direction finding, \ourprotocol achieves
 accurate real-time indoor localization using a few anchoring
 nodes  with known locations.
These anchoring speakers will emit acoustic signals using non-audible
 frequency (typically around 20kHz).
The smartphones play the role of receivers.
As it is difficult for a smartphone to find an accurate \emph{North}
 as base for absolute direction, our localization method does not 
 exploit the  absolute direction.
Instead we use a simple ``triangulation'' method by exploring the
 accurate opening angle from phone to two anchoring speakers.
 \ourprotocol let each phone measure the direction to source and its relative displacement for achieving  precise localization and  real-time tracking respectively. Anchor nodes will not perform any computation or communication.  Thus, \ourprotocol  supports
 \emph{arbitrary} number of users with extremely low cost.

We designed, deployed, and evaluated \ourprotocol for both direction
 finding and  real-time indoor localization.
Our extensive experimental results show that
 \ourprotocol  supports high accuracy for both direction finding and
 real-time indoor  localization.
In our testing of \ourprotocol, the finder only needs to shake the phone
 \emph{gently} and in \emph{arbitrary} patterns, which is  different
 from  the method in 
 \cite{2012-MobiQuitous2011-ProposalDirectionEstimation} as it requires
 the user to stretch the arm and then swing the phone through 180 degrees.
 For the phone-to-phone direction finding, the mean error of
 the measured angle is $2.10^o$ within the range of $32m$, and the
 errors are under $2.06^o$, $4.43^o$, $5.81^o$  at  50\%, 90\%, 95\%
 respectively, when  the acoustic source faces towards to the phone. 
For indoor localization, we deploy one acoustic source  per 6 meters,
 which broadcasts signals at a predefined frequency.
For indoor localization, \ourprotocol  achieves $90$-percentile
 accuracy of $0.92m$, maximum error of $1.73m$, and the mean error of
 $0.5m$.
For real-time indoor tracking, the error is always kept within $0.4m$
 even when users walk for  more than $50$ meters.

The rest of the paper is organized as follows.
We review the related work in Section~\ref{sec:related} and
 present technical preliminaries in Section~\ref{sec:pre}.
We present the acoustic direction finding of \ourprotocol in
 Section~\ref{sec:directionfinding},
 and indoor localization and tracking in
 Section~\ref{sec:localization}.
We report our extensive experiment
results in Section~\ref{sec:experiment}.
We conclude the paper in Section~\ref{sec:conclusion}.

\section{Related Work}
\label{sec:related}

\subsection{Direction Finding}

\noindent\textbf{Specialized Hardware:}  
One type of approaches is by using directional antenna
\cite{4711074,4509717,Niculescu:2004:VBS:1023720.1023727} or antenna array \cite{DBLP:journals/cee/KulakowskiVELG10} to implement Angle of Arrival (AOA) \cite{2003-INFOCOM-AdHocPositioning} in localization.  For
example, by rotating the beam of its antenna, a receiver can pinpoint
the direction of the AP as the direction that provides the highest
received strength \cite{4509717}.

\noindent\textbf{Non-specialized hardware:}
 \cite{2011-MOBICOM-Iamantenna} effectively emulates the
sensitivity and functionality of a directional antenna by rotating the
phone around the user's body, to locate outdoor APs. \cite{2011-SenSys-feasibilityrealtime}
leverages 2 microphones at each phone, \ie, at least 4 microphones,
for calculating 3D position of each other by using the distance
ranging method \cite{2007-SenSys-BeepBeephighaccuracy}. As the work is
intended for high-speed, locational, phone-to-phone (HLPP) games, it does not show the result when two phones are in long distances. Another
method \cite{2009-MobiSys-Point&Connectintentionbased} close to
direction finding is to identify which target the user is pointing at
when s/he moves mobile phone towards the target phone. 

To the best of our knowledge, the approach closest to ours in
direction finding is 
\cite{2012-MobiQuitous2011-ProposalDirectionEstimation}. It  
estimates the direction by using Doppler effect and achieves the mean angular errors within $18^o$. This approach
requires the searching user generates a Doppler Effect to all
directions, \eg, the user stretches the arm while holding the
searching device, and then swings it through 180 degrees. \ourprotocol only requires that the user shakes the phone in an
arbitrary path. 

\subsection{Indoor Localization and Tracking}
\noindent\textbf{Wireless Localization:} 
A significant advantage of wireless
localization is that it only leverages an existing infrastructure
instead of special-purpose hardware. Hence it attracts many research
efforts, \eg, \cite{Youssef:2005:HWL:1067170.1067193
,DBLP:conf/infocom/BahlP00,2010-MOBICOM-Didyousee,2012-MOBICOM-Locatinginfingerprint,2012-MOBICOM-PushlimitWiFi,2012-MOBICOM-Zeezeroeffort}.
However, it is found \cite{2012-MOBICOM-PushlimitWiFi} that the
wireless localization, such as the WiFi-based localization, can
achieve reasonable accuracy (\eg, $3\sim4m$), but there always exist
large errors (\eg, $6\sim8m$) unacceptable for many scenarios.  There
have  been many schemes proposed recently that improve the accuracy, such as using hundreds of APs \cite{5168931}, or adding
additional constraints by  
  exploiting the coordination among several phones 
 running this application in a small area \cite{2012-MOBICOM-PushlimitWiFi}.

\noindent\textbf{Infrastructure-based Localization:} 
There have been myriad approaches of indoor localization based on
special-purpose 
infrastructure. They are based on alternative signals, \eg,  
infrared \cite{Want:1992:ABL:128756.128759}, acoustic
\cite{626982}, visual
\cite{DBLP:journals/trob/SeLL05}. \textit{These approaches can
achieve high accuracy, but the need for special-purpose hardware and
infrastructure is a significant challenge} \cite{2012-MOBICOM-Zeezeroeffort}.
Cricket \cite{Priyantha:2000:CLS:345910.345917} uses concurrent radio
and ultrasonic signals to infer distance and obtain the location. 
ByteLight \cite{byteLight} claims to be able to provide low-price
infrastructure for localization using 
ceiling-embedded LEDs which send out Morse Code-like  signals
to be detected by the smartphone's camera. 

Our prototype provides another choice for precise indoor localization,
which only needs the off-the-shelf speakers,  or
even the loudspeakers installed in the mall, which can beep using high
frequency channel  without affecting normal broadcast. 

\noindent\textbf{Leveraging the acoustic wave by phone:} 
The methods of leveraging the acoustic wave in smartphone
applications have been well addressed. Most of them are leveraging the
low speed of the acoustic wave compared to wireless signals, such as
the mechanism of TOA \cite{2007-SenSys-BeepBeephighaccuracy} and TDOA \cite{DBLP:conf/mobicom/YangSCVLCCGM11}. BeepBeep
\cite{2007-SenSys-BeepBeephighaccuracy} detects the distance
 between two smartphones with  high accuracy. 
It has been used by many other schemes, such
 as HLPP games
\cite{2011-SenSys-feasibilityrealtime,2012-MobiSys-SwordFightenablingnew},
device pairing \cite{2009-MobiSys-Point&Connectintentionbased} and indoor localization
\cite{2012-MOBICOM-PushlimitWiFi,2012-MOBICOM-Centaurlocatingdevices}.

In this work, we leverage the Doppler effects of the acoustic waves
(\ie, measuring the precise relative displacement and velocity of phone) to
design \ourprotocol for direction finding and indoor localization.
\ourprotocol is precise enough to be another basic tool of AOA, while
it only requires off-the-shelf speakers.  Furthermore, \ourprotocol
supports arbitrary number of users and the phones of users do not
need to send any signals to get the location, which avoids the signal
interference when the number of users increases.

\noindent\textbf{Leveraging the Doppler effects:}
Doppler effects have been leveraged in wide areas, such as radar,
satellite communication, medical imaging and blood flow measurement,
etc. There are also localization approaches leveraging the Doppler
shift of wireless signals in localization
\cite{2008-SenSys-Spinningbeaconsprecise} and tracking
\cite{2007-SenSys-Trackingmobilenodes} in wireless sensor networks. But it also needs special hardware not available for smartphone users.
Meanwhile, by using the phase shift, \ourprotocol 
easily implements precise tracking without complicated algorithms
compared with \cite{ 2007-SenSys-Trackingmobilenodes} which uses
frequency shift.

\noindent\textbf{Leveraging the inertial sensors:} Inertial sensors
have been used for pedestrian dead-reckoning \cite{citeulike:7912602} in indoor
localization.  The challenge is that it suffers from large accumulation
of errors. The complementary
approaches to this problem are proposed in
\cite{Wang:2012:NNW:2307636.2307655, 2012-MOBICOM-Zeezeroeffort}.
\ourprotocol uses the accelerometer and gyroscope to obtain the
direction of the acoustic source.

\section{Preliminary Approaches}
\label{sec:pre}

\subsection{Mapping from Doppler Effects to Motion}
Our scheme is based on the relationship between Doppler effects and the
relative motion from the phone to the acoustic source, when the phone
moves and causes Doppler effects on the received acoustic waves.
Suppose the acoustic source is emitting the sinusoidal signal  at the frequency of $f_a$, the observed
frequency $f_r$  \cite{rosen2009encyclopedia} is
$f_{r}=\frac{v_{a}+v}{v_{a}+v_{s}}f_{a}$.
Here $v$ is the velocity of the receiver;
positive if the receiver is moving towards the source and negative in
the opposite position.  $v_s$ is the velocity of the source and $v_a$ is the traveling speed of the acoustic wave.

In this paper, we only consider the circumstance that the acoustic
source is motionless or the velocity of the phone is far greater than
the source, \ie, $v\gg v_s$.
As typically $v_a\gg v_s$,
we simplify the computing of the frequency shift $f$ as follows:
\begin{equation}
    f=f_r-f_a=\frac{v-v_{s}}{v_{a}+v_{s}}f_{a}\approx
    \frac{v}{v_{a}+v_{s}}f_{a} \approx
    \frac{f_a}{v_a}v\label{eq:doppler}
\end{equation}
We also assume the acoustic source sends the consecutive sinusoidal
acoustic wave at constant frequency $f_a$.
To  derive the relative displacement from Doppler effect, we  assume
that the received signal has the form:
\begin{equation}
    r(t)=A(t)\cos(2\pi f_{a}t+\phi(t))+\sigma(t)
\label{eq:received-signal}
\end{equation}
where $A(t)$ is the amplitude which changes continuously,
$\phi(t)$ is the phase which is affected by the Doppler effect and
$\sigma(t)$ is the noise. Assuming $\phi(t)$ is a continuous function,
the observed frequency $f_r$ at  time $t$ is
$    f_r(t)=\frac{1}{2\pi}\frac{d( 2\pi f_{a}t+\phi(t))}{dt}=
    f_a+\frac{1}{2\pi}\frac{d\phi(t)}{dt}.$
From \eqqref{eq:doppler}, the frequency shift $f$ at  time $t$ is
\begin{equation}
    f(t)=\frac{1}{2\pi}\frac{d\phi(t)}{dt}
    \label{eq:f(t)}
\end{equation}

From \eqqref{eq:doppler}\eqref{eq:f(t)},
we get the velocity and displacement relative to the acoustic source:
\begin{eqnarray}
    \begin{cases}
        v(t)=\frac{v_{a}}{2\pi f_{a}}\frac{d\phi(t)}{dt} \\
        s(t)=\frac{v_{a}}{2\pi f_{a}}\phi(t)-\frac{v_{a}}{2\pi f_{a}}\phi(0)
    \end{cases}
    \label{eq:v-and-phi}
    \label{eq:s-and-phi}
\end{eqnarray}
where $s(t)$ is the relative displacement from the phone to the acoustic source. Specifically, $s(t)=L(0)-L(t)$, where $L(t)$ is the distance between the phone and the source at
time $t$.
In Section~\ref{sec:stvt}, we further show how to calculate $\phi(t)$
in order to obtain $v(t)$ and $s(t)$.

\subsection{Basic Direction-Finding Using Doppler Effect for Simple Motion}

We make a simple case of phone-to-phone direction finding to
illustrate the intuition and challenges in designing \ourprotocol.
\begin{figure*}[t]
    \begin{center}
        \begin{subfigure}[t]{0.24\textwidth}
            \includegraphics[height=1.2in]{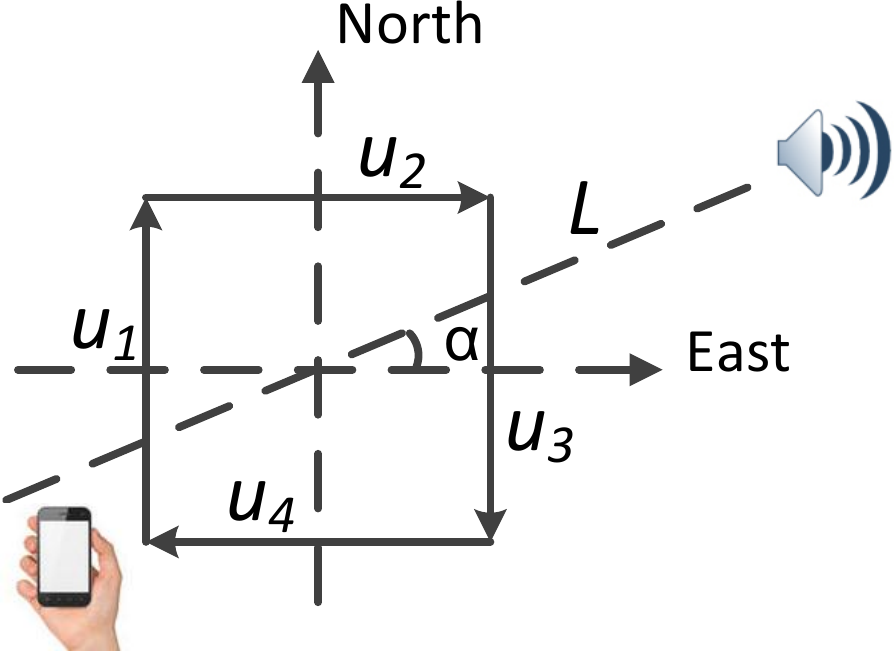}
            \caption{Phone movement.}
            \label{fig:simplemovement}
        \end{subfigure}
        \begin{subfigure}[t]{0.22\textwidth}
            \includegraphics[height=1.2in]{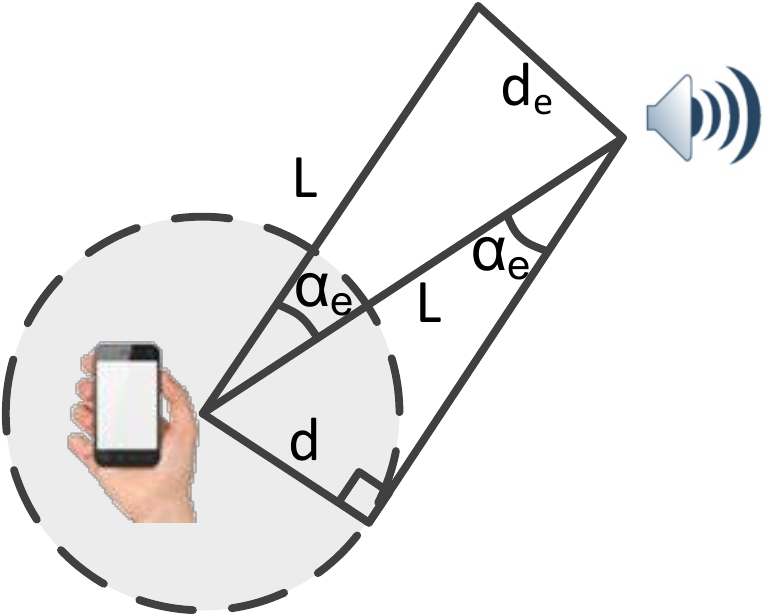}
            \caption{Error from motion.}
            \label{fig:directionerror}
        \end{subfigure}
        \begin{subfigure}[t]{0.2\textwidth}
            \includegraphics[height=1.2in]{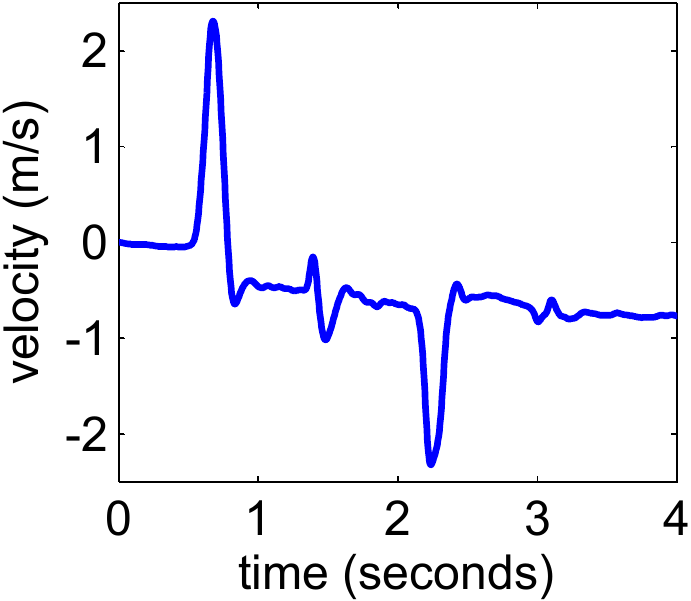}
            \caption{Velocity at north axis.}
            \label{fig:accumlateerror}
        \end{subfigure}
        \begin{subfigure}[t]{0.3\textwidth}
            \includegraphics[height=1.2in]{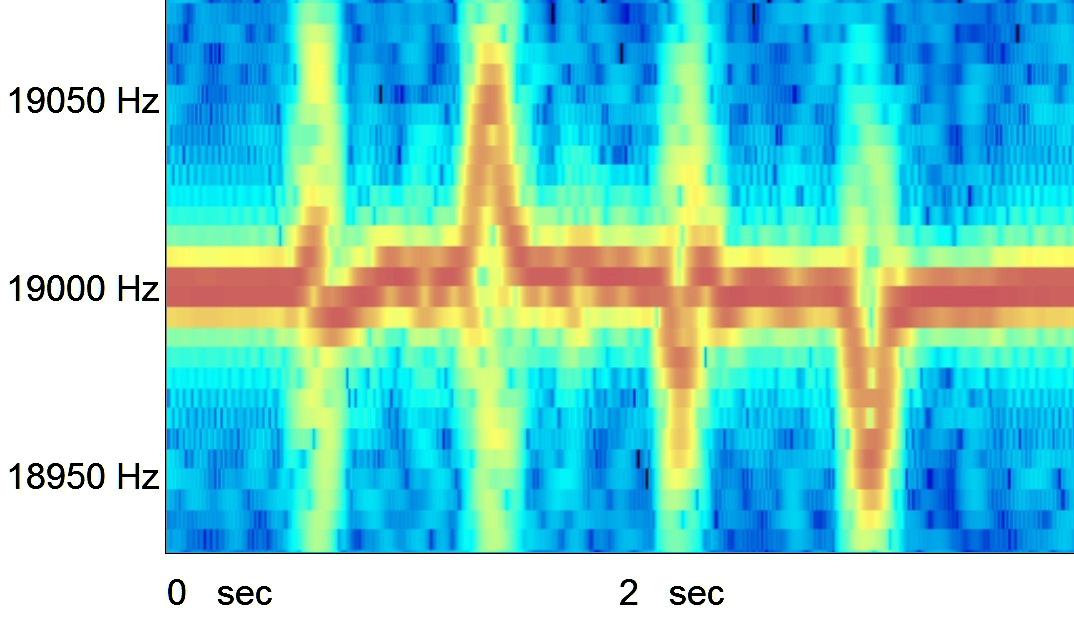}
            \caption{FFT of received signal.}
            \label{fig:fftresolution}
        \end{subfigure}
        \caption{A simple case of calculating the direction $\alpha$.
            (a) The phone starts moving  north and draw a rectangle. (b) The error $\alpha_e$ and $d_e$ caused by the movement of the phone. (c)
            The velocity calculated from the inertial sensors. (d) FFT on the
        received acoustic signal.}
        \label{fig:asimplecase}
    \end{center}
\end{figure*}

Assume that the phone and the acoustic source are
at the same height and the mobile phone starts moving in north and in
a path of rectangle with the constant velocity $u_{1}$, $u_{2}$,
$u_{3}$, $u_{4}$ in each direction, shown in Figure
\ref{fig:simplemovement}. So, frequency shifts are generated, where
$f_i$ corresponding to $u_i$.  If the velocities and the frequency  
shifts are obtained, from \eqqref{eq:doppler}, we can calculate the direction in the following
equations:
\begin{equation}
    \begin{cases}
        u_{1}\sin\alpha  =\frac{v_{a}}{f_{a}}f_{1}; \ \ \
        & u_{2}\cos\alpha  =\frac{v_{a}}{f_{a}}f_{2};\\
        -u_{3}\sin\alpha  =\frac{v_{a}}{f_{a}}f_{3}; \ \ \
        & -u_{4}\cos\alpha  =\frac{v_{a}}{f_{a}}f_{4}
    \end{cases}\label{eq:initialequation}
\end{equation}
Intuitively from \eqqref{eq:initialequation}, if $u_1=u_2=u_3=u_4$,
$f_2>f_1>0>f_3>f_4$,
which indicates that $0<\alpha<45^o$. Formally, only two equations
are needed to calculate $\alpha$ if the velocity in one equation is
not parallel to the other. The additional equations can improve the
accuracy by using maximum likelihood estimation.

Note that $\alpha$ is changing while the phone is moving, so it will
cause errors on calculating $\alpha$. However, it won't affect much
on calculating the direction. In Figure \ref{fig:directionerror}, if the initial distance from the
phone to acoustic source is $L$ and the maximum moving range of the
phone is $d$, the maximum angle error is
$    \alpha_e=\arcsin \frac{d}{L}.$
As the phone moves gently, we assume that $d$ is $10cm$ at maximum.
The maximum errors are $5.7^o$, $1.15^o$, $0.57^o$, $0.19^o$ at $L=1, 5, 10,
30m$ respectively, \ie,
the errors get smaller when the distance becomes longer.

Moreover, if the phone calculates
the position of acoustic source by not only the direction $\alpha$
according to \ourprotocol but also the distance $L$ according to other
techniques such as BeepBeep
\cite{2007-SenSys-BeepBeephighaccuracy} while the measured
$L$ is accurate, the distance $d_e$  from the calculated position to the actual position is
$ d_e=2L\sin\frac{\alpha_e}{2}=2L\sin\frac{\arcsin(d/L) }{2}$.
When $d\ll L$, \eg, $10d\leq L$, $\arcsin(d/L)\approx (d/L)$ and
$\sin(d/2L)\approx d/2L$. So we simplify $d_e$ as
$ d_e\approx d$.
Then the maximum error on computed location caused by shaking is 
close to the shaking distance $d$, which is acceptable in direction finding.

However, there are several problems on applying this simple approach.
First, the accurate velocity of the phone is hard to be  obtained by
using the inertial sensors. Though it can be calculated by the
accelerometer and other sensors if given the initial velocity of the
phone, the errors of the acceleration will be accumulated on its
integration, \ie, the calculated velocity. For instance in Figure
\ref{fig:simplemovement}, the velocity is zero at the end of moving
while the calculated one is $-0.77m/s$ in Figure \ref{fig:accumlateerror}. Second, the mobile
phone and the acoustic source may not be of the same height. In this
case, the calculated $f$ is lowered and the equations in
\eqqref{eq:initialequation} are not right. Third, it would be hard and
exhausting to draw the regular rectangle  for the
phone users. Fourth, the velocity of the phone $v$ cannot be constant in each
direction.
So we  need a more general solution in cases of different heights and
arbitrary motion patterns.

 Normally, the velocity increases and then decreases, as
shown in Figure \ref{fig:accumlateerror}. The rapid changes of $v$
bring the difficulties on calculating the frequency shift $f$.
Specifically, spectrum analysis, such as Fast Fourier Transform
(FFT), is efficient in calculating $f$, if $v$ is large or close to constant
for a while. But FFT cannot measure the precise value of $f$ if $v$
changes quickly due to the time-frequency resolution problem
\cite{claerbout1992earth}. That is, for any signal, the time duration
$\Delta T$ and the spectral bandwidth $\Delta F$ are related by
$\Delta F \Delta T \geq 1$.  For example, in Figure
\ref{fig:fftresolution}, we try to apply FFT on the received signal,
where the frequency of the acoustic wave is $f_a=19000$Hz, the sample
rate is 44100Hz, and FFT size is 8192. So, the time resolution is
$\Delta T=8192/44100$Hz$=0.19s$. Then, the frequency resolution
$\Delta F\geq 1/\Delta T=5.38$Hz. However, we assume that the maximum
speed of a user's hand is $2m/s$
\cite{2012-MobiSys-SwordFightenablingnew}.
The maximum frequency shift is
$f_{max}=2*19000/340=111.8$Hz. Even if the maximum speed is
satisfied, the relative velocity may not reach $2m/s$. For instance,
when the maximum speed of phone is
about $2m/s$ shown in Figure \ref{fig:accumlateerror}, for the phone
never moves towards directly to the acoustic source, the maximum
frequency shift is about $60$Hz in Figure \ref{fig:fftresolution},
which corresponds to the relative velocity $v=1.1m/s$.
Furthermore, in our circumstance, we only require that the user shakes the
phone gently, so most of the time the frequency shift is far less than
$111.8$Hz.
The resolution $\Delta F$, which is more than 5.38Hz, is not precise
enough to measure the frequency shift.

Hence, if the relative velocity and corresponding frequency shift are  
close to constant for a period, designers can increase $\Delta T$ to
get better frequency resolution by FFT. However, in our circumstance,
the velocity is always changing, which requires that both $\Delta T$
and $\Delta F$ is small enough, to get more precise $f$ at smaller
time block. Hence, it is in conflict with the time-frequency resolution
problem of FFT for estimating $f$.

Besides the challenge of calculating the frequency shift $f(t)$ for direction finding, the
further problem is calculating the phase shift $\phi(t)$, from which $f(t)$ can be obtained by \eqqref{eq:f(t)}. We also show that the real-time indoor tracking can be implemented by using $\phi(t)$ in Section \ref{sec:realtimetracking}.


\section{Acoustic Direction Finding}
\label{sec:directionfinding}

\begin{figure}[h]
\begin{centering}
\includegraphics[width=3.3in]{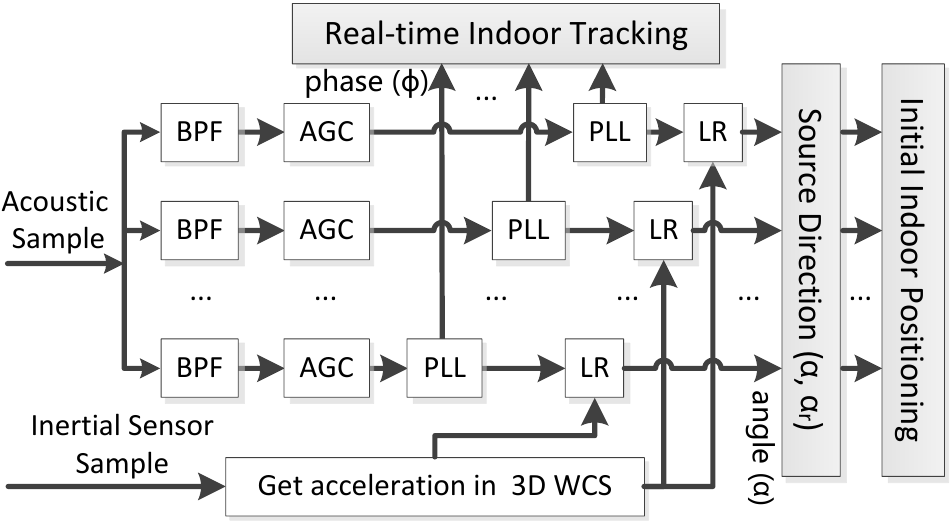}
\par\end{centering}
\caption{Implementation of \ourprotocol.}
\label{fig:Calculating-the-direction}
\end{figure}

In this section, we present the acoustic direction finding component
 of \ourprotocol.
We show the design of \ourprotocol in Figure
\ref{fig:Calculating-the-direction}.  The phone gathers samples from the
microphone, gyroscope and the accelerometer, when the user shakes
the phone or walks in an arbitrary path. The data are processed in real time to
maximize the utilization of the CPU. The phone dynamically updates the
direction of the source according to the previous calculated samples.

In Figure \ref{fig:Calculating-the-direction}, The noise $\sigma(t)$ and variational amplitude $A(t)$
in \eqqref{eq:received-signal} is eliminated by BPF and AGC
respectively. The \textit{phase} $\phi$ and \textit{frequency} $f$ are then
obtained by PLL. \ourprotocol further combines the velocity from the acoustic and
inertial sensor samples to get the \textit{source direction} $\alpha$ in LR.
The phone  returns the value of $\alpha$  and $\phi$ in real time for
\textit{direction finding}, \textit{indoor localization} or \textit{tracking}.  We describe each
component of the design as follows.

\subsection{Band Pass Filter (BPF)}
To get rid of the interference of other acoustic waves, we assume
the phones of different users send acoustic waves in different frequency bands.
Hence, in our implementation, the acoustic sample first walks through
the Band Pass Filter (BPF) such that only the waves at the specific
frequency pass through BPF. The interference by other acoustic sources
and the low frequency noises that human can hear are both eliminated.

Note that the type of BPF should be carefully
chosen. All frequency components of a signal are delayed when passed
through BPF. As the frequency is changing in Doppler effect and we
need to get the precise phase, the delay at each frequency components
must be constant, such that the different frequency component will not
suffer distortion, which is known as the linear phase property. As
a result, we choose equiripple FIR filter, which
satisfies the linear phase property.

Meanwhile, the bandwidth should
be wide enough to get the total signal.  Normally, the maximum speed
of shaking the phone is less than 2m/s.  Thus, if the frequency of
acoustic signal is $f_{a}=19000$Hz, the maximum frequency  
shift $f_{max}=111.8$Hz.  So, the minimum
pass band of the filter is 223.6Hz.  For avoiding the interference by
other acoustic sources, there should not be multiple signals that
pass through the same BPF.  Besides, acoustic bandwidth that the
almost all the smartphones support is limited with maximum of 22050Hz (\ie,
sample rates of 44100Hz) and we find that the lowest frequency that
human can hardly hear is about 17000Hz in our experiment.  Thus, the
maximum number of acoustic sources that can sound simultaneously in
a small area (with radius about $30m$) and be successfully detected
is limited to $(22050-17000)/223.6 \approx 23$. However, this is not
a challenge for \ourprotocol as we show that we only need a small
number (less than 10) of acoustic sources in a small area for high
accuracy.  Though there are possible ways to allow more
simultaneous acoustic waves such as dividing the signal into
different time slots, like TDMA in shared medium network, it is beyond the scope of this paper.

\subsection{Automatic Gain Control (AGC)}
We adjust the filtered data by Automatic Gain Control (AGC) such that
the amplitude of the acoustic signal $A(t)$ in \eqqref{eq:received-signal} is replaced by another one that is close to constant.
The purpose is to let the magnitude of $(\theta[k+1]-\theta[k])$ in
\eqqref{eq:adaptive} only be determined by $\mu$, rather than $A(t)$,  which is discussed in Section \ref{sec:stvt}.
We adopt the design of AGC from \cite{rice2008digital}.  Suppose $T_{s}$ is the sampling period of the received signal and  $k$ is the step count of sampling, then $t=kT_{s}$. The main
idea is for the  input $r_{b}[k]$ from BPF, we estimate the amplitude
$A[k]$ in \eqqref{eq:received-signal} by updating $A_{1}[k]$ with
the equation:
\begin{equation*}
\log(A_{1}[k])=(1-A_\alpha)\log(A_{1}[k-1])-A_\alpha\log(A_{r}[k-1])
\end{equation*}
Here $A_\alpha$ represents the sensitivity for adjusting $A_{1}[k]$. $A_{r}[k]$
represents the coarse-grained estimation of $A[k]$. In our implementation,
 $A_{r}[k]={\textstyle \frac{1}{7}\sum_{i=k-10}^{k}|r_{b}[i]|}$ and $A_\alpha=0.9$.
Then, for the received filter data $r_b[k]$, the output
\begin{equation*}
r_{c}[k]=A_{1}[k]r_{b}[k]
\end{equation*}

For the amplitude of $r_{c}[k]$ is close to constant by AGC, if
$A_{1}[k]=A_{1}[k-1]$, $A_{1}[k]A_{r}[k-1]=1$. Thus, the amplitude
of $r_{c}[k]$ is close to 1.
Hence, we get $r_c(t)\approx \cos(2\pi f_a t+\phi(t))$, where
$\sigma(t)$ and $A(t)$ in \eqqref{eq:received-signal} is eliminated by BPF and AGC respectively.

\subsection{Phase Locked Loop (PLL)}
\label{sec:stvt}
According to \eqqref{eq:v-and-phi}, we use Phase
Locked Loops (PLL) to calculate the phase $\phi(t)$, in order to get
the precise relative displacement $s(t)$ and velocity $v(t)$ of the phone.
PLL can be thought as a device that tracks the phase and frequency of
a sinusoid \cite{rice2008digital}. In software implementation, we draw
the idea from \cite{citeulike:5657344}.
To get the precise $\phi(t)$, we update an adaptive estimation of
$\phi(t)$ in real time, denoted as
$\theta(t)$ in order that $\theta(t)\approx \phi(t)$.  To make
$\theta$ converge to $\phi$ after enough iterations, we define the
corresponding function $J_{\textit{PLL}}(\theta)$ such that
$J_{\textit{PLL}}$ converges to its maximum at the same time.
Specifically, $\theta(t)$ is updated in the iterations as:
\begin{equation}
    \theta'=\theta+\frac{dJ_{\textit{PLL}}}{d\theta}
        \label{initialadapt}
\end{equation}
As a result, $J_{\textit{PLL}}$ should satisfy that
\begin{equation}
 \max (J_{\textit{PLL}}(\theta))=J_{\textit{PLL}}(\phi)
 \label{JPLLsatis}
\end{equation}

In \ourprotocol, we choose $J_{\textit{PLL}}$ as follows:
\begin{eqnarray*}
    J_{\textit{PLL}}(\theta) & = & \textrm{LPF}\{r_c(t)\cos(2\pi f_{a}t+\theta(t))\}\nonumber \\
\MYCUT{ 
& = & \frac{1}{2}\textrm{LPF}\{A(t)\cos(\phi(t)-\theta(t))\nonumber \\
 &  & \ \ \ \ +A(t)\cos(4\pi f_{a}t+\theta(t)+\phi(t))\nonumber \\
 &  & \ \ \ \ +2\sigma(t)\cos(2\pi f_{a}t+\theta(t))\}\nonumber \\
 & = & \frac{1}{2}\textrm{LPF}\{A(t)\cos(\phi(t)-\theta(t))\}\nonumber \\
 &  & \ \ \ \ +\frac{1}{2}\textrm{LPF}\{A(t)\cos(4\pi f_{a}t+\theta(t)+\phi(t))\}\nonumber \\
 &  & \ \ \ \ +\textrm{LPF}\{\sigma(t)\cos(2\pi
  f_{a}t+\theta(t))\}\nonumber \\} 
 & \approx & \frac{1}{2}\textrm{LPF}\{\cos(\phi(t)-\theta(t))\}
\end{eqnarray*}
Here, LPF is the Low Pass Filter which excludes the high frequency component in the above approximation.
Hence, $J_{\textit{PLL}}$ satisfies \eqqref{JPLLsatis}.

Next,  we need to change the continuous estimation process of \eqqref{initialadapt} to the discrete
one.
Assuming a small step size, the derivation in \eqqref{initialadapt}
with respect to $\theta$ at $kT_{s}$ can be approximated\footnote{The
  proof of the approximation is in G.13 of \cite{citeulike:5657344}.}:
\begin{eqnarray*}
    \frac{dJ_{PLL}}{d\theta} & \approx & \left. \textrm{LPF}\{\frac{d[r_c[k]\cos(2\pi f_{a}kT_{s}+\theta))]}{d\theta}\} \right|_{\theta=\theta[k]}\\
    & = & -\textrm{LPF}\{r_c[k]\sin(2\pi f_{a}kT_{s}+\theta[k])\}
\end{eqnarray*}

As a result, the estimating of $\theta(t)$ is shown as follows:
\begin{equation}
    \theta[k+1]=\theta[k]-\mu \textrm{LPF}\{r_c[k]\sin(2\pi
        f_{a}kT_{s}+\theta[k])\}\label{eq:adaptive}
\end{equation}
where $\theta[k]=\theta(kT_{s})$ and $\mu$ is a small positive value.
Hence, $\phi[k]\approx\theta[k]$ after enough iterations.
According to \eqqref{eq:v-and-phi}, if the max velocity of the phone is
$v_{max}=2m/s$,
$f_{s}=$44100Hz and $f_{a}=19000$Hz, the max offset per sample
$|\Delta\phi_{max}|=\frac{2\pi f_{a}}{v_{a}f_{s}}v_{max}=0.016$.
Besides,
\[
\begin{array}{l}
     r_{c}[k]\sin(2\pi f_{a}kT_{s}+\theta[k]) \approx  \frac{1}{2}\sin(4\pi f_{a}kT_{s}+2\theta[k])\leq \frac{1}{2}
\end{array}
\]

Thus, $\mu>0.03$
in \eqqref{eq:adaptive}, otherwise, the transition rate of $\theta[k]$
cannot catch up with the real phase. Furthermore, as
$\frac{1}{2}\sin(4\pi f_{a}kT_{s}+2\theta[k])$ cannot always be $1/2$,
$\mu$ needs to be much more than $0.03$ to let $\theta[k]$ converge
to $\phi[k]$.
However, when $\mu$ is bigger, the calculated phase is more sensitive
to noises, and cannot be precise either. Hence, there is a trade off on
choosing the $\mu$.  In the implementation, we choose $\mu=0.03$.

\subsection{Leveraging Sensors}
The acceleration in world
 coordinate system (WCS) is calculated by using accelerometer and
 gyroscope of the phone.
As compass is not accurate,  we make the following implementation to
 avoid the error of compass.
The accelerometer records the 3D acceleration in user's phone
 coordinate system (UCS).
So, we convert the acceleration in UCS to the one in WCS as follows: 1) On
initialization, by leveraging the force of gravity of the earth
\cite{zaxis}, the Z axis in WCS is calculated by the accelerometer.
Typically Z axis is accurate.
The X axis in WCS is computed from the values of compass and gyroscope,
 which is supposed to point to the east but often has large errors due
 to noisy data.
2) After initialization, the conversion
 function is updated by using the gyroscope.

\begin{figure}[h]
\begin{center}
\begin{subfigure}[b]{0.18\textwidth}
\includegraphics[height=1.2in]{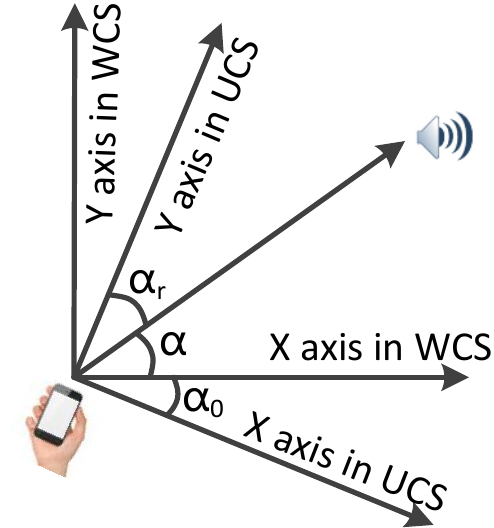}
        \caption{WCS vs. UCS}
\end{subfigure}
\begin{subfigure}[b]{0.23\textwidth}
\includegraphics[height=1.2in]{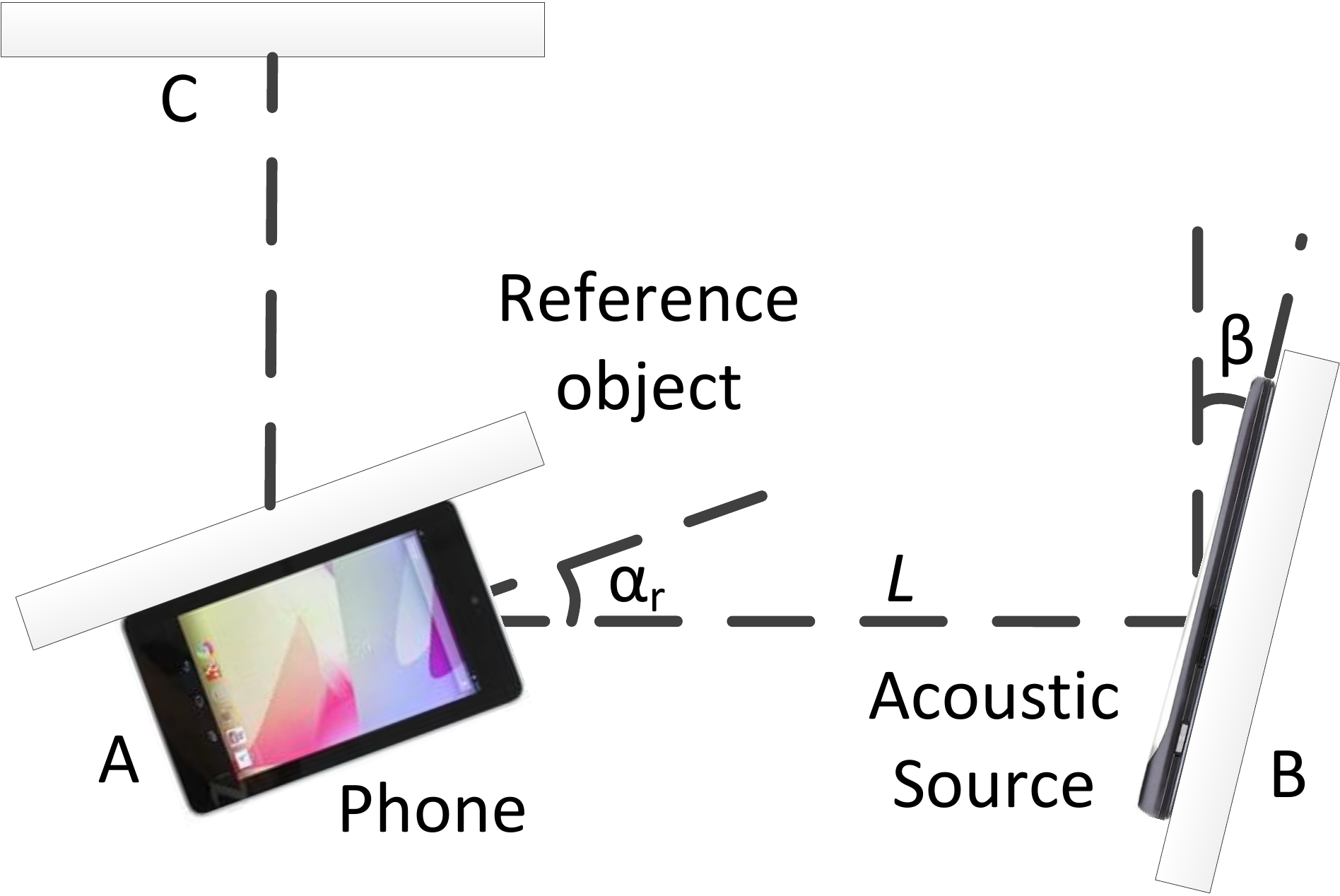}
        \caption{Experiment setup}
\end{subfigure}

    \caption{(a) WCS vs. UCS when the phone is horizontal. (b) Experiment of direction finding.}
    \label{fig:duallayout}
\end{center}
\end{figure}

Hence in our WCS, the Z axis is  considered to be accurate, but the X
 axis may not point to east.
So, the calculated direction $\alpha$ in WCS may not be the actual
 direction relative  to the east.
To evaluate the performance of our direction finding,
 we will evaluate the direction (denoted as $\alpha_r$) of the
 acoustic source using the UCS
 of the phone that is placed horizontally such that its Z axis is same
 as the Z axis of WCS, as shown in
 Figure  \ref{fig:duallayout}a.
When phone is static, the value $\alpha_r$ does not change.
Thus, in Section~\ref{sec:p2pexp}, we measure $\alpha_r$ to evaluate the
 precision of direction finding shown in Figure \ref{fig:duallayout}b.

Hence, suppose the phone is horizontal, we
get value $\alpha$ by using \ourprotocol  and the opening angle from X axis in UCS to the one in WCS ($\alpha_0$) by using the
transform function from UCS to WCS. $\alpha_r$ is calculated by
\begin{equation}
\alpha_r=\pi/2-\alpha-\alpha_0
    \label{alpha_r}
\end{equation}

\subsection{Getting Direction by Linear Regression (LR)}

\MYCUT{
In 3D world's coordinate system, where the X axis points toward East,
Y axis points to the North Pole, and Z axis points to the sky. In
this case, $\alpha$ is the direction in the projected 2D world's
coordinate system. }
Assuming the direction vector of the acoustic
source relative to the phone is
$\overrightarrow{\lambda}=(\lambda_{x}, \lambda_{y}, \lambda_{z})$ and
velocity vector of the phone is $\overrightarrow{u}=(v_{x}, v_{y},
v_{z})$, then
$\overrightarrow{u}\cdot\overrightarrow{\lambda}=\frac{v_{a}}{f_{a}}f$
according to \eqqref{eq:doppler}.
For the obtained array $\overrightarrow{u}[k]$ and $f[k]$, they
satisfy the following equations
\begin{equation}
    \lambda_xv_x[k]+\lambda_yv_y[k]+\lambda_zv_z[k]=\frac{v_{a}}{f_{a}}\cdot
    f[k], \quad \forall k
    \label{eq:linearinitial}
\end{equation}
Hence, the 3D direction $\overrightarrow\lambda$ can be obtained by
solving these equations  using linear regression,
where $f[k]$ can be calculated by \eqqref{eq:f(t)}, \eqqref{eq:adaptive}.
Ideally, if $u[k]$ is obtained from inertial sensors and
there are no errors of $u[k]$, there are 3 unknowns
$\lambda_x,\lambda_y,\lambda_z$ in the equation set.
Moreover, using this we can calculate the direction
 when  the phone moves in arbitrary paths,
 because different motion patterns of the phone
 merely causes different array
 $\overrightarrow{u}[k]$ and $f[k]$.

We can also translate 3D direction $\overrightarrow\lambda$ to 2D direction $\alpha$ as follows:
\begin{equation}
\alpha=\begin{cases}
    \arcsin\frac{\lambda_{y}}{\sqrt {\lambda_{x}^2+\lambda_y^2}} & \lambda_x \geq 0\\ \pi+\arcsin\frac{\lambda_{y}}{\sqrt {\lambda_{x}^2+\lambda_y^2}} & \lambda_x<0 \end{cases}
\label{eq:arctan} \end{equation}

 \begin{figure*}
     \begin{centering}
         \begin{subfigure}{0.30\textwidth}
             {\includegraphics[height=1in]{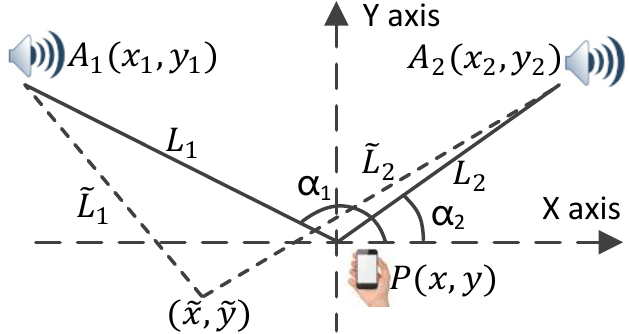}}
             \caption{Trilateration}
             \label{fig:Fine-grained-Indoor-Location}
         \end{subfigure}
         \begin{subfigure}{0.16\textwidth}
             {\includegraphics[height=1in]{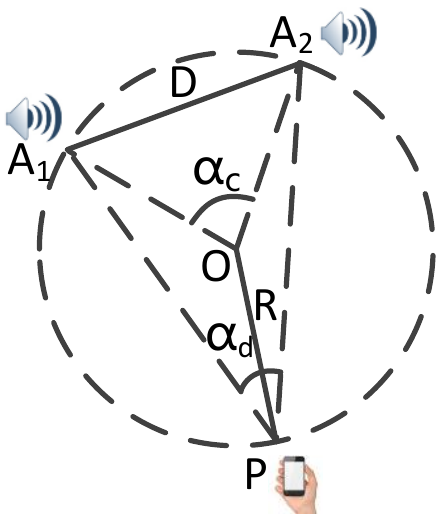}}
             \caption{Acute Angle}
             \label{fig:solveindoor-a}
         \end{subfigure}
         \begin{subfigure}{0.16\textwidth}
             {\includegraphics[height=1in]{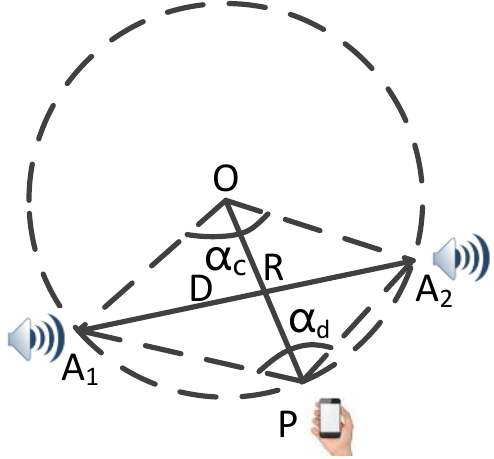}}
             \caption{Obtuse Angle}
             \label{fig:solveindoor-b}
         \end{subfigure}
         \begin{subfigure}{0.16\textwidth}
             {\includegraphics[height=1in]{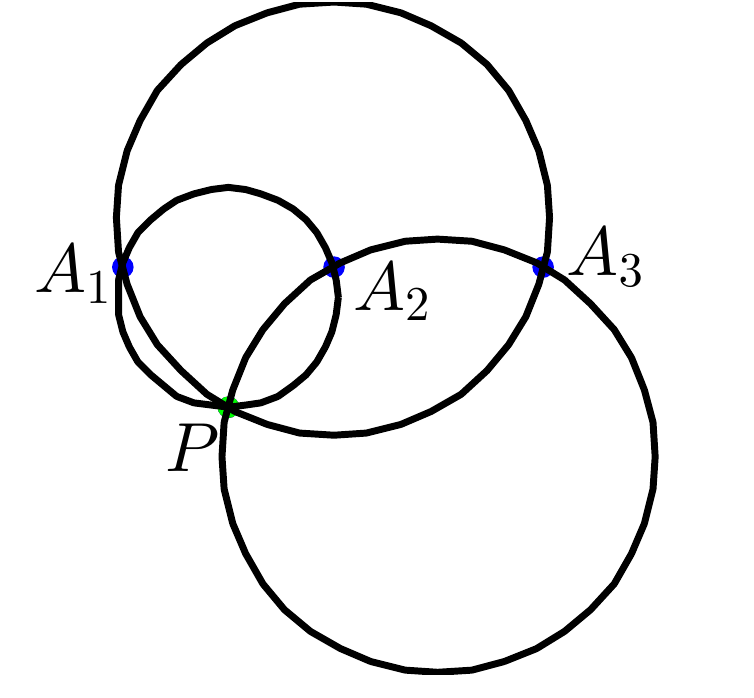}}
             \caption{Good layout}
             \label{fig:layout-g}
         \end{subfigure}
         \begin{subfigure}{0.18\textwidth}
             {\includegraphics[height=1in]{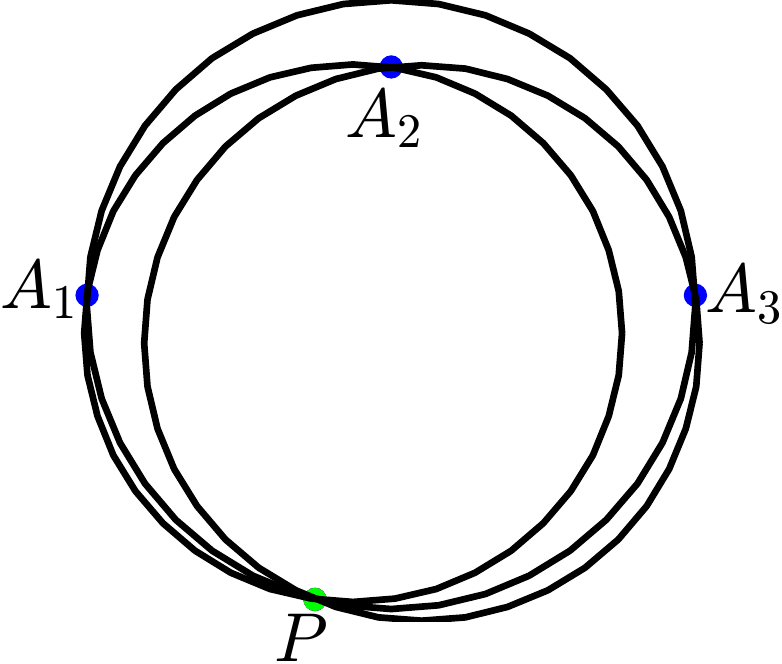}}
             \caption{Bad layout}
             \label{fig:layout-b}
         \end{subfigure}
         \caption{Indoor localization and tracking: trilateration, pinpoint
             candidate location to a circle (acute angle and obtuse angle), and
         impact of layout  of anchors (good and bad).}
         \label{fig:location-cases}
     \end{centering}
 \end{figure*}

We now address non-ideal circumstance with noisy sensor data, \ie,
 to minimize the error of velocity
 which is derived from the calculated acceleration in WCS.
In phone-to-phone direction finding and indoor localization, we only need
the 2D direction $\alpha$ rather than the 3D direction
$(\lambda_x,\lambda_y,\lambda_z)$. Thus, $\lambda_z$ is not needed. From \eqqref{eq:linearinitial}, if
$\lambda_zv[k]\approx 0$, \ie, the phone moves in a horizontal plane or the two phones are at
the same height approximately, we can calculate the direction by
 the following equation to eliminate the error of $v_z$:
\begin{equation}
    \lambda_xv_x[k]+\lambda_yv_y[k]=\frac{v_{a}}{f_{a}}\cdot f[k]
    \label{eq:simplydirection}
\end{equation}

Suppose $\hat{a}_x[i]=a_x[i]+\sigma_x[i]$ where $\hat{a}_x[i]$, $a_x[i]$,
$\sigma_x[i]$ is the real acceleration, the calculated acceleration,
the error of the calculation on the acceleration of the $i$th sample respectively. We can
derive $v_x$ from
\[
v_x[k]=v_x[0]+\sum_{i=0}^{k-1}T[i]a_x[i]+\sum_{i=0}^{k-1}T[i]\sigma_x[i]
\]
where $T[i]$ is the time interval from $a_x[i]$ to $a_x[i+1]$.

The error $\sigma_x$ is related the natural quality of the inertial
sensors and challenging to be measured. In this paper, we simply
assume $\sigma_x$ equals to a constant $e_x$ at a short period.
Suppose $t[k]=\sum_{i=0}^{k-1}T[i]$, we get
$\sum_{i=0}^{k-1}T[i]\sigma_x[i]=e_xt[k]$. Similarly, we also assume
the error of $a_y$ is a constant $e_y$ at a short period.

As a result, from
Eq. \eqref{alpha_r}\eqref{eq:arctan}\eqref{eq:simplydirection}, we
could calculate the 2D direction by linear regression from the following
equation set which has 4 unknowns ($\lambda_x$, $\lambda_y$,
$\lambda_0$, $\lambda_1$)
\begin{equation*}
\left(\begin{array}{cccc}
    w_{x}[0] & w_{y}[0] & 1 & t[0] \\
    w_{x}[1] & w_{y}[1] & 1 & t[1]\\
    \cdots & \cdots & \cdots & \cdots \\
    w_{x}[n] & w_{y}[n] & 1 & t[n]
\end{array}\right)\left(\begin{array}{c}
\lambda_{x}\\
\lambda_{y}\\
\lambda_{0}\\
\lambda_{1}
\end{array}\right)=\frac{v_{a}}{f_{a}}\cdot\left(\begin{array}{c}
f[0]\\
f[1]\\
\cdots\\
f[n]
\end{array}\right)
\end{equation*}
where $w_x[k]=\sum_{i=0}^{k-1}T[i]a_x[i]$,
$w_y[k]=\sum_{i=0}^{k-1}T[i]a_y[i]$,
$\lambda_0=\lambda_xv_x[0]+\lambda_yv_y[0]$ and
$\lambda_1=\lambda_xe_x+\lambda_ye_y$. Note that, we allow that
$v_x[0]\not=0$ and $v_y[0]\not=0$ in our solution, which means we
don't require the phone to be motionless before shaking the phone and
calculating the direction. $v_x[0]$ and $v_y[0]$ are put together as an unknown
$\lambda_0$ in the equation.

\section{Indoor Localization \& Tracking}
\label{sec:localization}

We now describe our basic
 method in \ourprotocol for  fine-grained indoor localization
 illustrated in Figure \ref{fig:Fine-grained-Indoor-Location}, which is based on the direction $\alpha$ and the phase $\phi$ in Section \ref{sec:directionfinding}.
We require that there are at least three acoustic sources as
 anchor nodes installed, which send sinusoid signals at
 the specific different  frequencies.
Users need to get the position and frequency of
 each anchor node from network service.
\ourprotocol includes two phases:
 finding the initial position and real-time tracking.

\subsection{Finding the initial position}
The user needs to shake the phone first in order to get his/her initial
position.
The phone calculates the direction of each
anchor node in WCS and then gets the position.
Note that as the compass is not precise, the calculated directions,
 such as $\alpha_1$, $\alpha_2$ in
 Figure~\ref{fig:Fine-grained-Indoor-Location}, are not directly  used in
 calculating the  position.
However,  observe that the opening angle
 $(\alpha_1-\alpha_2)$ is fixed no matter which WCS is chosen.
We calculate the initial position using this opening angle.
\MYCUT{
\begin{figure}[htpb]
    \begin{center}
\includegraphics[scale=0.7]{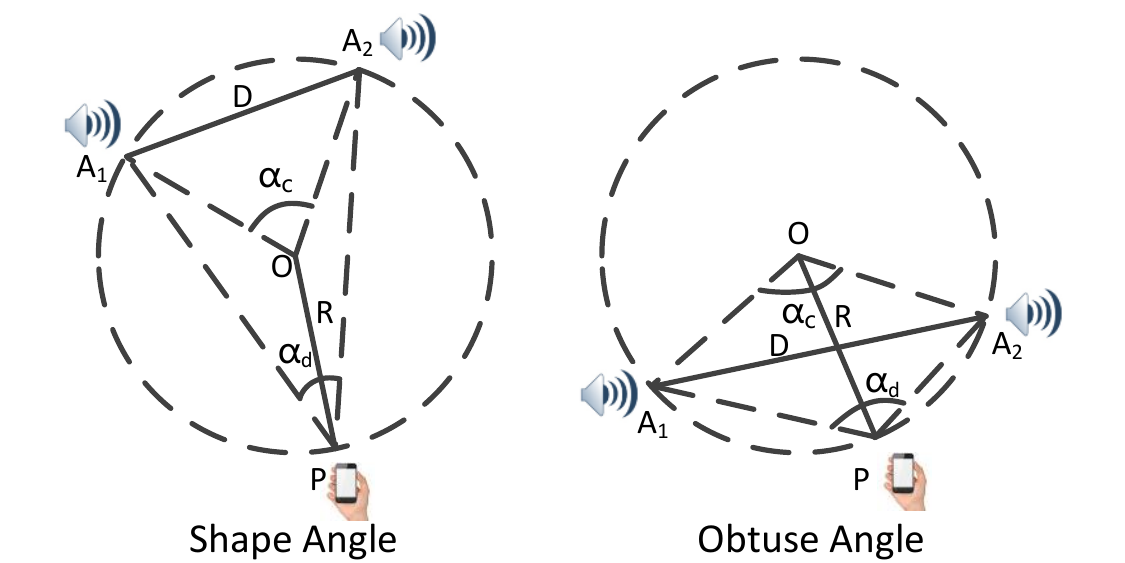}
    \end{center}
\caption{Getting the radius $R$ and the center of circle $O$ by using
  $D$ and $\alpha_d$.  }\label{fig:solveindoor}
\end{figure}
}
Taking the positions $(x_1, y_1)$ and $(x_2, y_2)$ of two anchor nodes
 $A_1$ and $A_2$ and the relative directions $PA_1$, $PA_2$ from phone (with unknown
 position $P$) to $A_1$ and $A_2$, we can compute the distance
 $D=\|A_1 -A_2\|$ and the opening angle $\alpha_d = \angle A_1 P
 A_2$, as illustrated in Figure~\ref{fig:Fine-grained-Indoor-Location}.
It can be inferred that the position $P$ is on a fixed circle
 illustrated in Figure~\ref{fig:solveindoor-a},~\ref{fig:solveindoor-b}.
If $\alpha_d$ is a cute angle as in Figure \ref{fig:solveindoor-a},
 $\alpha_c=2\alpha_d$.
So, the radius of the circle $R=\frac{D}{2\sin \alpha_d}$.
Then we get at most two possible solutions of the position
 of the circumcenter $O$ by using radius $R$ and the given coordinates
 of   two nodes $A_1$ and $A_2$.
If $\alpha_d$ is a cute angle, then $O$ and $P$ are on the same side
 of $A_1A_2$.
Similarly, if $\alpha_d$ is an obtuse angle, as in Figure \ref{fig:solveindoor-b},
 $O$ and $P$ are on the opposite side of $A_1A_2$.

For a system of $n$ anchor nodes, there are $\frac {n(n-1)}{2}$  pairs
of anchor nodes.
As a result, phone $P$ lies on $\frac {n(n-1)}{2}$ circles.
Thus, with at least 3 anchor nodes, we can get the position of $P$.
It is worth mentioning that for the circle formed by a node pair, the
circle is divided into two arcs by the node pair.
Node $P$ only lies on one of the arcs,
 depending on whether $\alpha_d$ is an acute angle or an obtuse angle.
Hence, for localization we search for the point $P$
 to minimize  $ \sum_i d_i$ where $d_i$ is the distance
 from $P$ to the $i$th arc.

We claim that it will result in better
localization accuracy if we place the anchor nodes in a line as in
Figure~\ref{fig:layout-g} compared to the one in
Figure~\ref{fig:layout-b}.
In Figure \ref{fig:layout-b}, the centers of the
 circles are too close, which causes big potential errors.
The root reason is that the 4 points $A_1$, $A_2$, $A_3$, $P$ are nearly at the
 same circle, which means the arbitrary point, \eg, $A_1$, is close to
 the circle which is constructed by the rest of 3 points, \eg, $A_2$, $A_3$, $P$.
\MYCUT{
\begin{figure}[htb]
\begin{subfigure}[b]{0.22\textwidth}
                \centering
\includegraphics[scale=0.4]{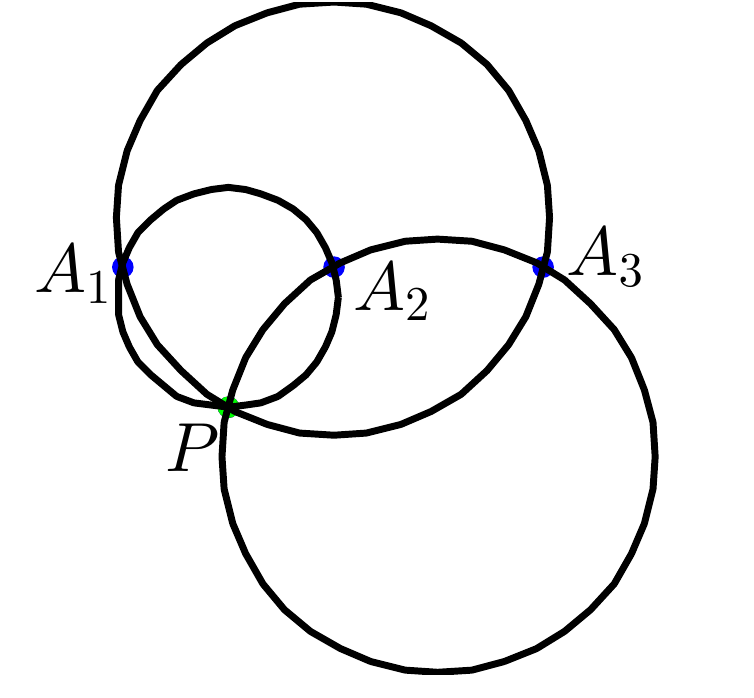}
                \caption{Good layout}
        \end{subfigure}
\begin{subfigure}[b]{0.2\textwidth}
                \centering
\includegraphics[scale=0.4]{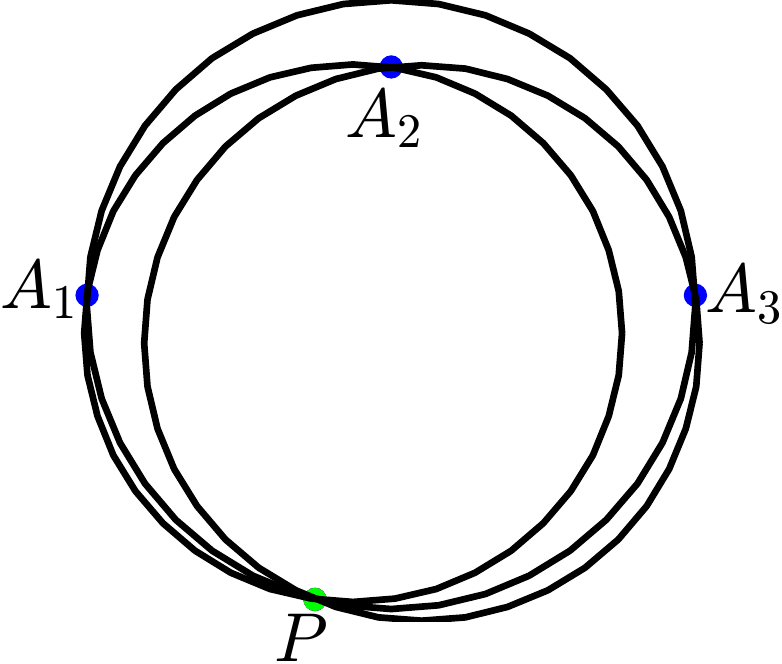}
                \caption{Bad layout}
        \end{subfigure}
\caption{The effect by the layout of the anchor nodes.
    \label{fig:layout}}
\end{figure}
}

\subsection{Real-time tracking}
\label{sec:realtimetracking}
After getting the initial location of phone, the phone then gets the
 real-time location  by  calculating the
 relative displacement to each anchor node without shaking the phone
 again.
In Figure \ref{fig:Fine-grained-Indoor-Location}, if the location of
 phone at time $t$ has been calculated,
 denoted as $(x,y)$, we calculate its location $(\tilde{x},\tilde{y})$
 at the latter time $\tilde{t}$ by getting $s(t)$ and $s(\tilde{t})$
 using \eqqref{eq:s-and-phi}, \eqqref{eq:adaptive}. 
Then we calculate next location according to
 $(\tilde{x},\tilde{y})$ iteratively.
Specifically,
 if the user gets the location $(x,y)$, then the distance from $(x,y)$
 to $(x_i,y_i)$ is
 $    L_{i}=\sqrt{(x-x_{i})^{2}+(y-y_{i})^{2}+h_i^2}$,
where $h_i$ is the relative height between the phone and the source
$(x_i,y_i)$. Thus, s/he gets the distances from all the available
acoustic sources at time $t$.
According to \eqqref{eq:s-and-phi}, \eqqref{eq:adaptive} and the definition of $s_i$, we have
\begin{equation}
        \tilde{L_i}=L_i-\frac{v_{a}}{2\pi f_{a}} (\tilde{\phi_i}-\phi_i)
\label{distance_in_specific_moment}
\end{equation}
where $\tilde{L_i}=L_i(\tilde t)$ and $\tilde \phi_i = \phi_i(\tilde t)$. 
Then we search for location
$(\tilde{x},\tilde{y})$ near $(x,y)$ to minimize $\sum _i M_i$ where
$    M_i=\big | \tilde{L_i}-\sqrt{(\tilde{x}-x_i)^2+(\tilde{y}-y_i)^2+h_i^2} \big|$.

\section{Experiment}
\label{sec:experiment}

We implement \ourprotocol on Nexus 7, where all the components, including BPF and PLL, are implemented by using Android APIs.
The audio sample rate is  44100Hz, and sample rate of the gyroscope and
 accelerometer is 200Hz.

\subsection{Phone-to-phone Direction Finding}
\label{sec:p2pexp}
\subsubsection{Experiment Design}

The vertical view of the phone and acoustic source is shown in Figure
\ref{fig:duallayout}b. The distance between the phone and the acoustic
source is $L$. The orientation angle of the phone and acoustic source
at the horizontal plane is $\alpha_r$ and $\beta$ respectively. There
are reference objects at places A, B, C which are used to align
the phones. The place C is used to put new acoustic source for
further experiment. Additionally, we assume elevation angle of the
acoustic source is $\gamma$ which is not shown in this 2D figure. The
acoustic source is  on the floor, the height of phone from the
floor is about $40cm$.

The main process of evaluating performance of direction finding is as
follows: we
vary $L$, $\alpha_r$, $\beta$, $\gamma$ by moving the reference
objects. We obtain the measured direction $\alpha_r$ by shaking
the phone, aligning the phone to the reference object, and reading
the direction value from the phone. We measure $\alpha_r$ 50 times
for each configuration.

\subsubsection{Empty Room with Single Acoustic Wave}

We first conduct the experiment in a large empty room  for
 examining the accuracy of direction finding
 when there is only single acoustic wave. The sound pressure of the room is $-41$ dBFS (about 30 dB SPL) measured by Nexus 7.  The amplitude of the acoustic source at the distance of $1m$ is $-20$ dBFS.

\begin{figure}[h]
\begin{subfigure}[b]{0.235\textwidth}
\includegraphics[height=1.5in]{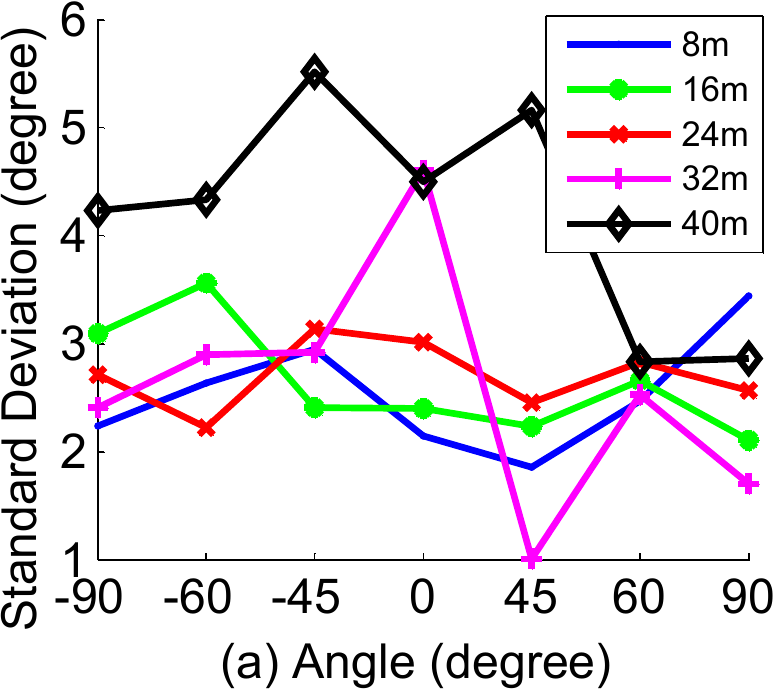}
        \end{subfigure}
\begin{subfigure}[b]{0.2\textwidth}
\includegraphics[height=1.5in]{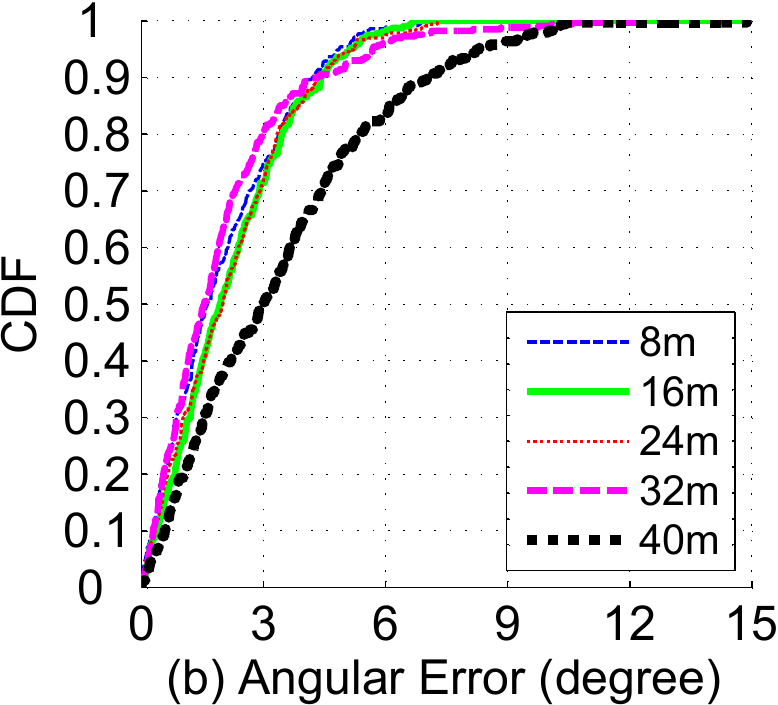}
        \end{subfigure}
    \caption{The result of direction finding in an empty room when $\beta=0$ and $\gamma=0$.}
    \label{fig:initresult}
\end{figure}

\noindent\textbf{Effect by $L$ and $\alpha_r$.} The cases we mostly
care about is the performance when the distance $L$ and the
orientation of the phone $\alpha_r$ is changing. Hence, we set
$\beta=0$ and $\gamma=0$, and plot the standard deviations
and cumulative distribution function (CDF) of the angular errors when $L$ and
$\alpha_r$ are changed in Figure \ref{fig:initresult}.

\begin{figure}[h]
    \begin{center}
        \includegraphics[width=3.35in]{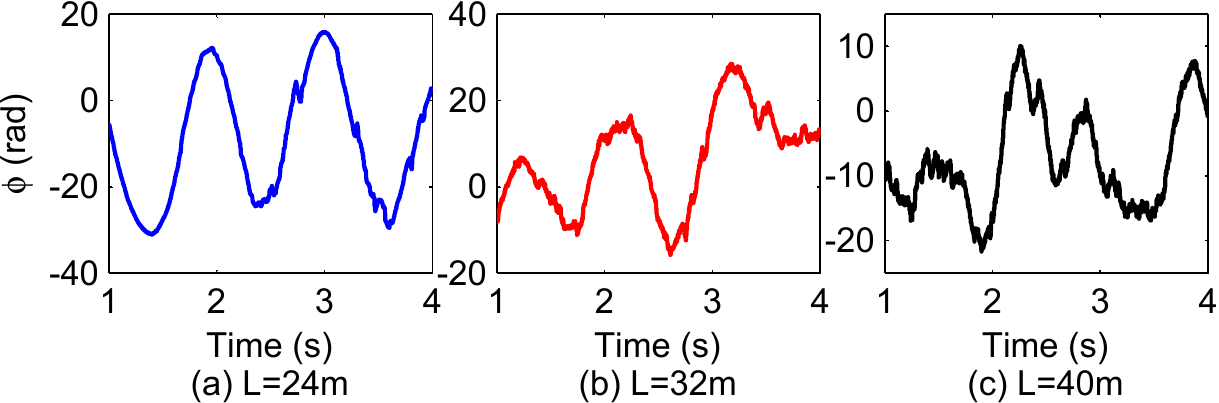}
    \end{center}
    \caption{The calculated phase $\phi(t)$.}
    \label{fig:phasemeter}
\end{figure}

The key observation is that the measurement is very precise when
$L\leq32m$.  We examine the
reason in Figure \ref{fig:phasemeter}, which plots the calculated
$\phi(t)$ on random samples with different $L$ values. The calculated
$\phi(t)$ is always smooth when $L \leq 24m$, while there are
small noises when $L=32m$ and much bigger noises if
$L=40m$. Hence, the calculated related displacement and velocity
become much less precise when $L=40m$, which affects the
calculation of direction. It is similar that most of the following
cases mainly affect the calculated phase which finally affect the
precision of direction finding.

\begin{figure*}[htpb]
    \begin{center}
        \includegraphics[width=7in]{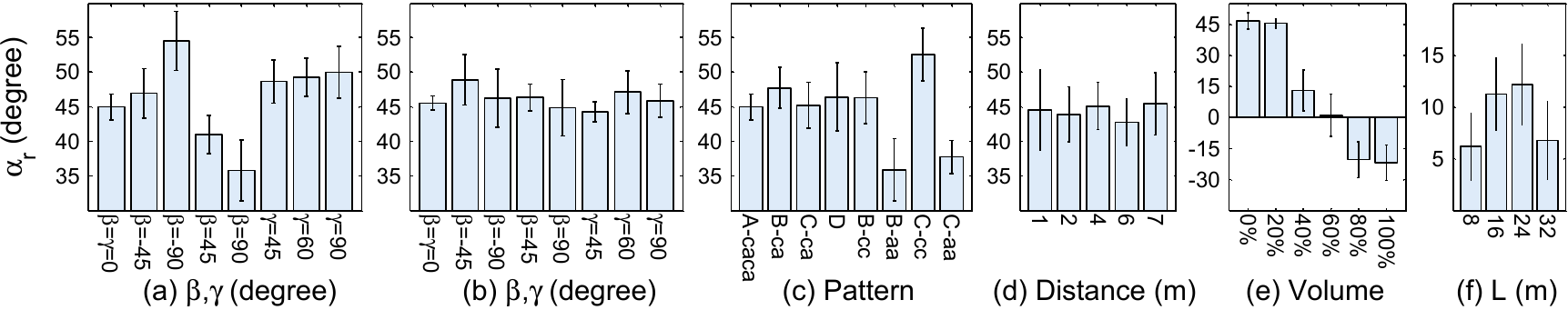}
    \end{center}
    \caption{Effect by (a) $\beta$ and $\gamma$ when $L=8m$
      (b) $\beta$ and $\gamma$ when $L=32m$ (c) motion
      pattern (d) non-line of sight (e) man-made multipath (f)
      multipath from the wall.}
    \label{fig:errorerror}
\end{figure*}
When $L\leq 32m$, the mean error and standard
deviation of the measurement is $2.10^o$ and $2.66^o$. The angular
errors are within $2.06^o$, $4.43^o$, $5.81^o$ at  50\%,
90\%, 95\% respectively. Though the errors become larger when
$L=40m$, it is still acceptable. We also test angle
errors when $L>40m$, but it becomes much unstable as the signal is
too weak. So we do not show the result of this case.

We also find that  $\alpha_r$ has little effect on precision according
to Figure~\ref{fig:initresult}a.  As the errors are so close for
different $\alpha_r$, we don't show the CDF of different $\alpha_r$.

\noindent\textbf{Effect by $\beta$ and $\gamma$.}
We test the errors when the orientation of the acoustic source is not
directly pointing to the phone. In this case, we set
$\alpha_r=45^o$. In Figure \ref{fig:errorerror}a, \ref{fig:errorerror}b, we show
the mean and standard deviation with different choices of
$\beta$, $\gamma$, $L$.

It shows an interesting result that when $\beta$
 changes, the mean value changes more in $L=8m$ than the one
 in $L=32m$. The main reason is that the acoustic source we
 choose is not omnidirectional, and the signal is much stronger right
 in front of the source.
The signal reflected from the wall affects the result, which is
 so-called the multipath effect.
When the phone is further from the source, the signal reflected
 from the wall becomes much weaker than the one directly from the acoustic source.

Another observation is that if the phone turns up, such as
 $\gamma=45^o$, $60^o$, $90^o $, the mean value will not
 change a lot no matter $L=8m$ or $L=32m$. That is, though there is
 multipath from the ceiling, it has little effect on the mean
 direction.
We find a new phenomenon on multipath effect in latter
experiment, which  explains these observations here.

\noindent\textbf{Motion Pattern.} We also analyze the angular errors
caused by the inertial sensors.
As we claim that \ourprotocol supports arbitrary pattern of phone
movement, we test errors caused by different motion patterns of the
phone. In this case, we set $L=32m$, $\alpha_r=45^o$,
$\beta=\gamma=0$.

\begin{figure}[h]
\begin{centering}
        \includegraphics[scale=0.7]{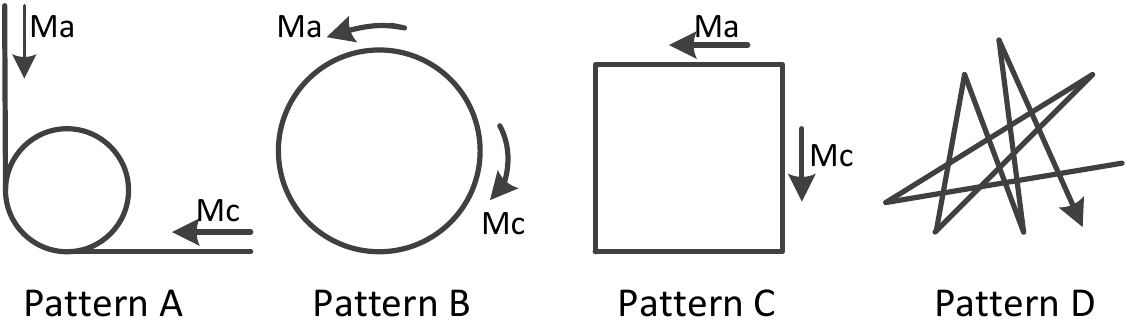}
\par\end{centering}
\caption{Basic phone motion patterns.}
    \label{fig:motionpattern}
\end{figure}

We define several motion patterns in Figure \ref{fig:motionpattern}.
Pattern A is the default basic pattern used in the whole experiment.
The pattern A is a mix of rectangle and circle. The pattern
B, C, D is the circle, the rectangle, and the arbitrary
pattern respectively. We shake the phone with the basic patterns
anti-clockwise or 
clockwise for a few times and get the result in Figure
\ref{fig:errorerror}c. The first motion pattern of this figure,
named \emph{A-caca}, means we shake the phone 4 times in basic pattern
A: \textbf{c}lockwise, \textbf{a}nticlockwise, \textbf{c}lockwise,
\textbf{a}nticlockwise. The rest of the patterns can be explained
similarly.

First of all, we found the result of arbitrary pattern D is still
 acceptable in $L=32m$: the standard deviation of the
 measurement is $4.96^o$.
Another important observation is that, when the phone moves clockwise,
there is a positive shift on the mean value. 
When the phone moves anti-clockwise, there is a negative shift. 
For the pattern D, there are both positive and negative shifts in the
 measurement, so the standard deviation becomes a little bigger.
We also observed that when the phone was shaken in other regular
 patterns compared to pattern D, the standard deviation
 becomes smaller.
That is, the error shift is close
 to constant in these cases.
We also find that when we shake the phone in A-caca, C-ca, the
means are close to same.
We leave it as a future work to understand why the phenomena happen.

\noindent\textbf{Non-line of sight.}
 We set $L=8m$, $\alpha_r=45^o$, $\beta=\gamma=0$, and
 test a simple case on the effect by Non-line of sight (NLOS).  In
 Figure \ref{fig:errorerror}d, a person stands between the phone
 and acoustic source, and we measure the errors related to the
 distance from the person to the phone.
It becomes apparent that when the person
 stands in either ends, the standard deviation is enlarged, while the person stands
 in the middle, it is close to the one without
 obstruction.
Hence, the person has little effect on
 direction finding, as long as  s/he is not too close to the acoustic
 source or the receiver.
This is also verified in the experiment of noisy environment. 

Another case of NLOS is that the user put his back to the source. The
signal turns so weak and the result becomes unstable. In this case, the user can
turn around to get the precise direction. The other possible
complementory method is to let user rotate the phone around the user's
body, similar to \cite{2011-MOBICOM-Iamantenna}.

\noindent\textbf{Multipath effect.} As the multipath effect is hard to
measure exactly, we first make a man-made multipath to find its
impact. Then, we make a simple real case to verify our finding.

We set $L=8m$, $\alpha_r=45^o$, $\beta=\gamma=0$ and add another
 phone as acoustic source placed at position C in
 Figure \ref{fig:duallayout}b.
The new source is also 8 meters from the phone. It beeps at the same
frequency with the source at B. The volume of the
source at B is constant 60\%.  We change the volume of the source at
C from 0\% to 100\%, and plot the Figure \ref{fig:errorerror}e.
When the volume is less than 20\%, it has little effect: the
standard deviation is low, and the mean value is slightly lowered.
There is an interesting phenomenon that when the volume becomes
larger, the angle becomes lower which is close to the direction of the
new source. However, the standard deviation becomes bigger when both
sources have high volume.

We then conduct experiment with both acoustic source and phone near the
wall. The wall is on the right hand side of the user while shaking the
phone.  We set $\alpha_r=\beta=\gamma=0$ and $L=8,16,24,32m$.
The result is shown in Figure \ref{fig:errorerror}f. $\alpha_r$
becomes bigger for all the distances which can be inferred from the
above conclusion. It can also be inferred that the strengths of
the reflected signals relative to the respective direct signals are different at each $L$, which causes different
mean shifts of $\alpha_r$.  The other observation is that the standard
deviation is low for each distance. Hence, reflected signal is weak
compared to the one directly from the acoustic source.

\subsubsection{Empty Room with Multiple Acoustic Waves}

To validate the robustness of \ourprotocol,
 we conduct two types of experiments: (1) an acoustic source broadcasts multiple
 signals at different frequencies, (2) multiple sources
broadcast signals at different frequencies.

\begin{figure}[htb]
    \begin{subfigure}[c]{0.235\textwidth}
       \includegraphics[height=1.57in]{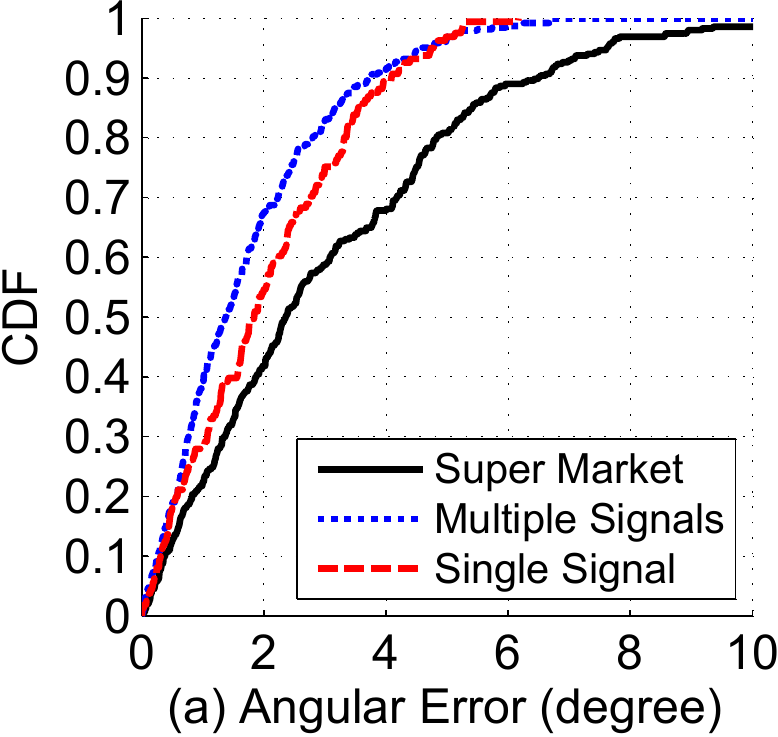}
        \end{subfigure}
    \begin{subfigure}[c]{0.22\textwidth}
       \includegraphics[height=1.57in]{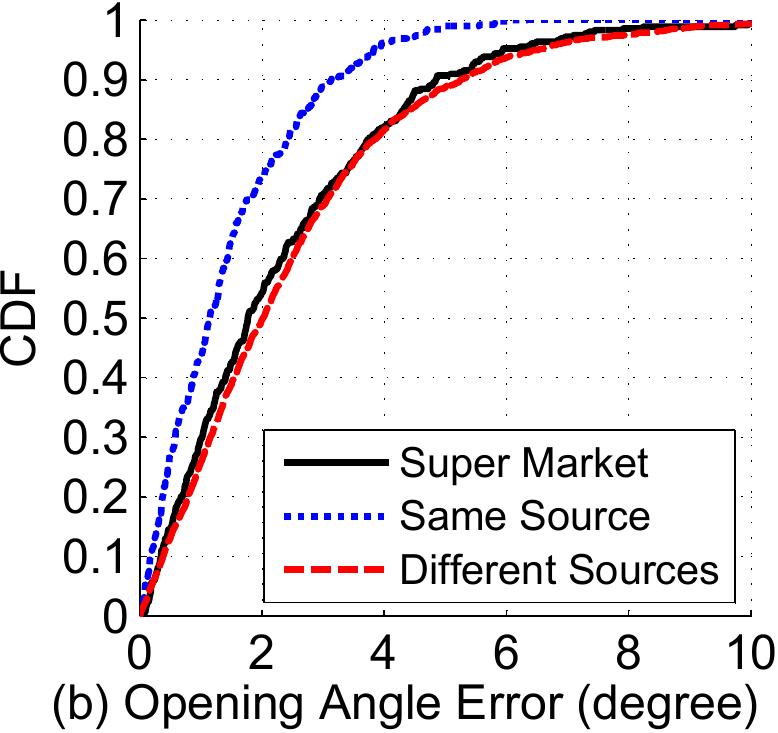}
        \end{subfigure}
        \caption{(a) Errors on different cases when $L\leq
            24m$, $\alpha_r=\beta=\gamma=0$. (b) The opening angle 
        errors when there are multiple signals.}
    \label{fig:errorandinnererror}
\end{figure}

In experiment (1), we measure the angular errors when 
the acoustic source sends 6 sinusoidal signals at the frequency from
17000Hz to 19500Hz. The experiment is performed by setting $\alpha_r=\beta=\gamma=0$. We find that the results are similar for different $L$ that $L\leq 24m$, while the ones at $L=32m$ are a little 
worse. It is
because that when the phone sends multiple signals, the signal
strength of each component becomes weaker.
We plot the CDF at $L\leq24m$ in Figure
 \ref{fig:errorandinnererror}a. The performance is almost the same with the one sending single
wave. It can be inferred that we can use
loudspeakers in the mall as anchor nodes while they are 
playing music.


We now  analyze
 the performance of direction finding when there are multiple acoustic
 sources.
The performance in this case will have direct impact on the accuracy
of the localization to be studied later in
Subsection~\ref{subsubsec:localization}.
Recall that as the computing of the absolute direction requires 
   the accurate compass which is hard to get,
  in our localization method we use the opening angle $\angle A_i P A_j$
   from the phone with location $P$ to two arbitrary anchor nodes
   $A_i$ and $A_j$ instead of the absolute orientation of any vector
   $PA_i$ or $PA_j$.
Thus, here we measure the accuracy of estimated angle $\angle A_i P
 A_j$ by varying the locations of $P$, $A_i$, and $A_j$.

Figure \ref{fig:errorandinnererror}b shows the opening angle
errors in three cases: (1) single source, multiple waves, super market, (2)
single source, multiple waves, empty room, (3) multiple source,
multiple waves, empty room.
We find that the opening angle  errors in
 cases (1), (2) are less than the direction errors in
 Figure~\ref{fig:errorandinnererror}a.
Furthermore, we observe that case (3) is much worse
 than (2).
Though it is unfair to compare the two cases that the
 acoustic sources are different, it shows the possibility of  improvement on the
 precision of indoor localization by using better acoustic sources, as
 we use the worse case for  calculating the latter position.

\subsubsection{Noisy Environment}
We conduct this experiment in a super market, where it is noisy ($-21$ dBFS) and there
 are people walking around and blocking the line from the acoustic source to the
 phone.
We also let the phone send multiple signals.  In Figure \ref{fig:errorandinnererror},
the result becomes a little worse than the one in empty room. Almost
all errors are less than 10 degrees, which is acceptable.

\subsubsection{Overhead}
As \ourprotocol calculates the direction in real time, we only
evaluate the CPU usage. When \ourprotocol processes
one acoustic signal, the CPU usage is 20.5\%. When it process
6 signals at the same time, the CPU usage of this application is
95.25\% and it takes the phone
3.9 seconds to process 1 second of signal samples on average. The main cost
for computation is the Band Pass Filter (BPF). We choose the FIR filter to
achieve linear phase property as discussed earlier. However, the
computation overhead is much higher than IIR filters. When there are
multiple signals, we need to shorten the bandwidth of the filter,
which costs more computation overhead. So there is a trade-off between
processing speed and accuracy:  we can enhance the speed by using IIR filter by sacrificing
a little accuracy. In fact, as we only need to shake the phone for a short duration to
get the directions, the overhead is not the key problem.

\begin{figure*}[htpb]
    \begin{subfigure}[c]{0.19\textwidth}
        \includegraphics[height=1.4in]{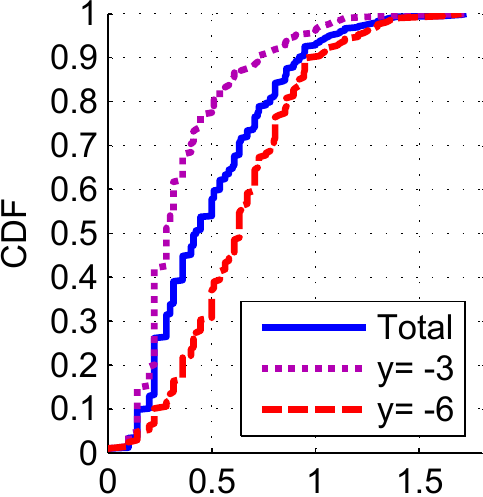}
        \caption{Position Error (m)}
    \label{fig:indoorCDF}
    \end{subfigure}
    \begin{subfigure}[c]{0.19\textwidth}
        \includegraphics[height=1.4in]{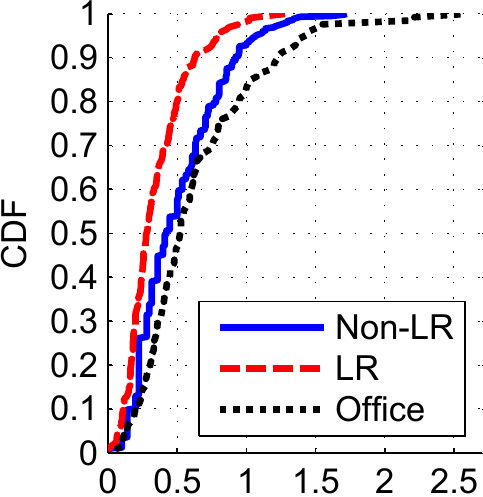}
        \caption{Position Error (m)}
    \label{fig:indoorCDFoffice}
    \end{subfigure}
    \begin{subfigure}[c]{0.19\textwidth}
        \includegraphics[height=1.4in]{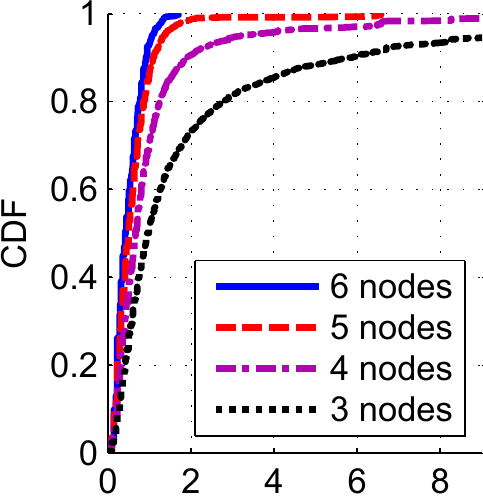}
    \caption{Position Error (m)}
    \label{fig:errordeletenode}
    \end{subfigure}
    \begin{subfigure}[c]{0.19\textwidth}
        \includegraphics[height=1.4in]{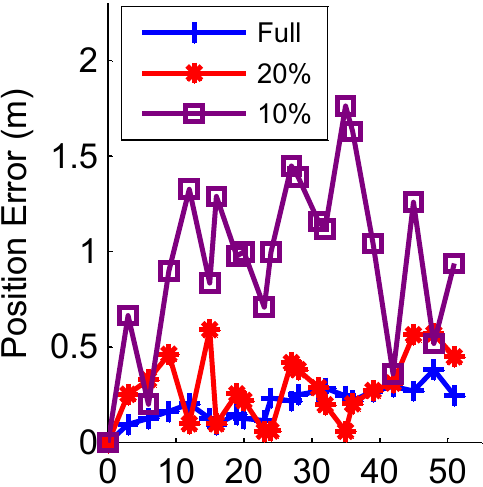}
    \caption{Walking Length (m)}
    \label{fig:trackingerrordetail}
    \end{subfigure}
    \begin{subfigure}[c]{0.19\textwidth}
       \includegraphics[height=1.4in]{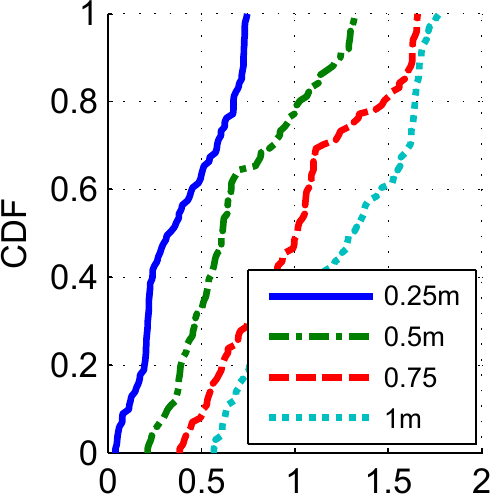}
        \caption{Position Error (m)}
        \label{fig:initialerroreffect}
        \end{subfigure}
    \caption{ Static localization accuracy (a) in different locations,
      (b) in different scenes and by different methods, (c) when parts
      of the anchor nodes are chosen for calculation.  Tracking
      accuracy (d) along the walking paths, (e) at final point
      $(24,-3)$ when there are initial position errors at $(6,-6)$.  } 
\end{figure*}

\subsection{Real-time Indoor Localization}
\label{sec:indorrexp}
\subsubsection{Experimental setup}

\begin{figure}[htb]
\begin{center}
\begin{tabular}{cc}
\includegraphics[width=1.5in,height=0.8in]{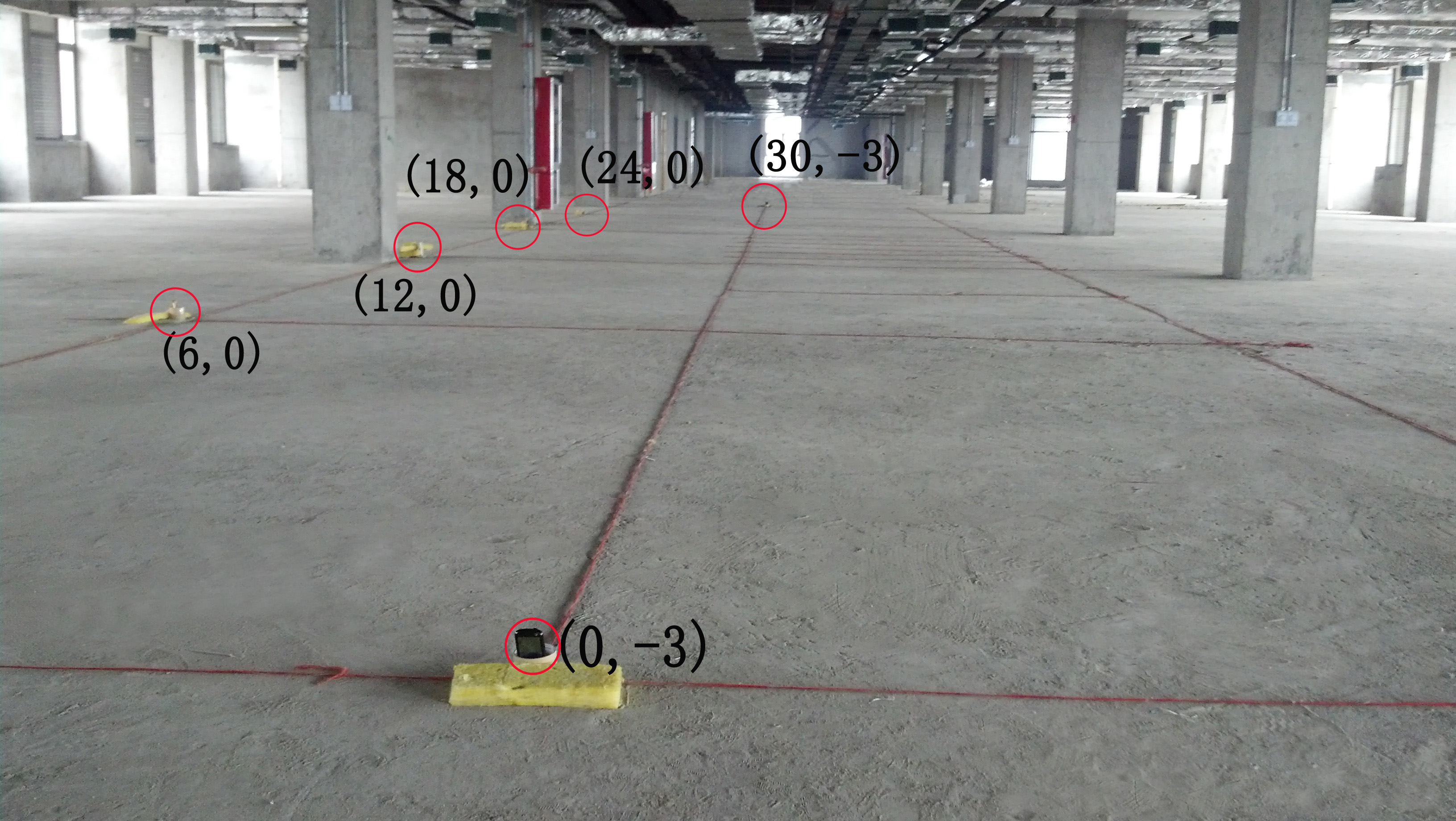}\quad &
\includegraphics[width=1.5in,height=0.8in]{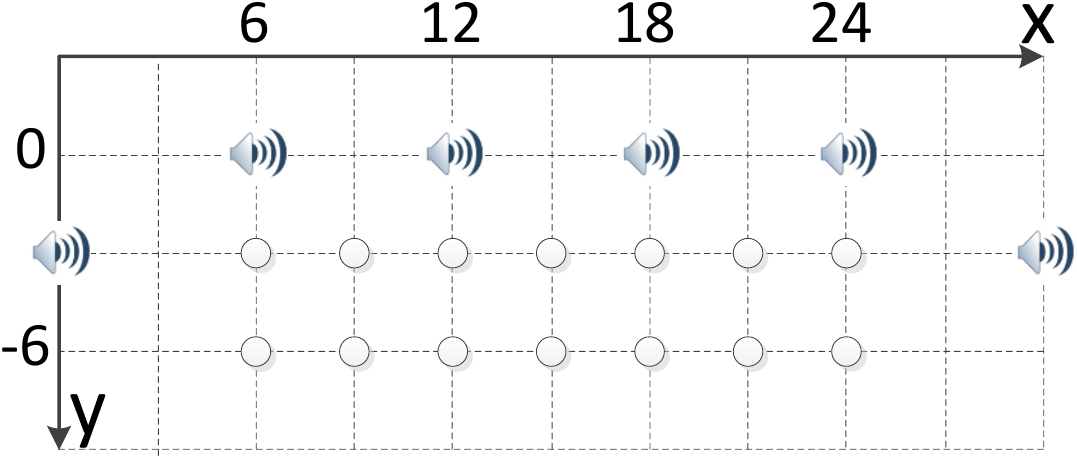}       \\
(a) Indoor environment & (b) Layout of anchors
\end{tabular}
\end{center}
\caption{Indoor localization testing prototype.}
\label{fig:indoorplacement}
\end{figure}

In Figure \ref{fig:indoorplacement}, we place 6 phones as anchor nodes
in the same empty room in the previous subsection. The positions are
$(0, -3)$, $(6, 0)$, $(12, 0)$, $(18, 0)$, $(24, 0)$, $(30, -3)$
(meters) respectively.  The beep frequencies are from 17000 to
19500Hz. We choose spots at $y\in \{-3, -6\}$ and $x\in
\{6,9,12, 15, 18, 21,24\}$.
We conduct the  localization when people stay at these spots, and
repeat the experiment  30 times for each spot.
How to place anchor nodes in optimal way in an area is left for future
research.

\subsubsection{Static Position Localization}
\label{subsubsec:localization}

The accuracy of  static  localization is shown in Figure
\ref{fig:indoorCDF}.
\ourprotocol achieves localization errors within $0.42m$,
$0.92m$, $1.08m$, $1.73m$ at the percentage of 50\%, 90\%, 95\%, and 100\%
respectively. The mean error and the standard deviation is $0.50m$ and
$0.59m$ respectively.
We also find that the localization accuracy at spots with $y=-3m$
 is better than the ones on $y=-6m$.
Specifically, on $y=-3m$,
 the localization errors are within $0.28m$, $0.73m$, $0.91m$, $1.73m$ at the
 percentage of 50\%, 90\%, 95\%, and 100\% respectively. 
 
Meanwhile, we find that there are nearly constant error shifts of the
 calculated position at all  locations. 
Thus, we further adjust the position by linear regression. 
That is, we build a polynomial function model from the calculated
 positions to more precise positions by learning the results from half 
 of the samples. 
We then apply the function to the other half and the result is ploted in Figure
\ref{fig:indoorCDFoffice}.
It shows that the 
 precision is greatly enhanced (\ie, the errors are within $0.67m$,
 $0.82m$, $1.56m$ at the 
 percentage of 90\%, 95\%, 100\% respectively). 
 
We then measure the errors of static localization
 in a large office (-34 dBFS), where the
 environment is much more complicated. 
The layout of the anchor nodes is nearly the same with the one in Figure
 \ref{fig:indoorplacement}, except the anchor nodes are installed on
 the ceiling. Figure \ref{fig:indoorCDFoffice} shows that the
 error is within $0.94m$, $1.23m$, $2.59m$ at the percentage of 80\%,
 90\%, 100\% respectively after linear regression.

 We also choose specific number of nodes (\ie, $3\sim 6$) from the
 6 nodes to calculate the position. In Figure
 \ref{fig:errordeletenode}, it shows that the precision is greatly
 enhanced when the number of nodes increases. Besides, the precision in
 case of 3 nodes becomes much worse for it is more sensitive by
 the layout shown in Figure  \ref{fig:layout-g},  \ref{fig:layout-b}.

\subsubsection{Real-time Tracking}

\begin{figure}[ht]
    \begin{center}
        \includegraphics[width=3.3in]{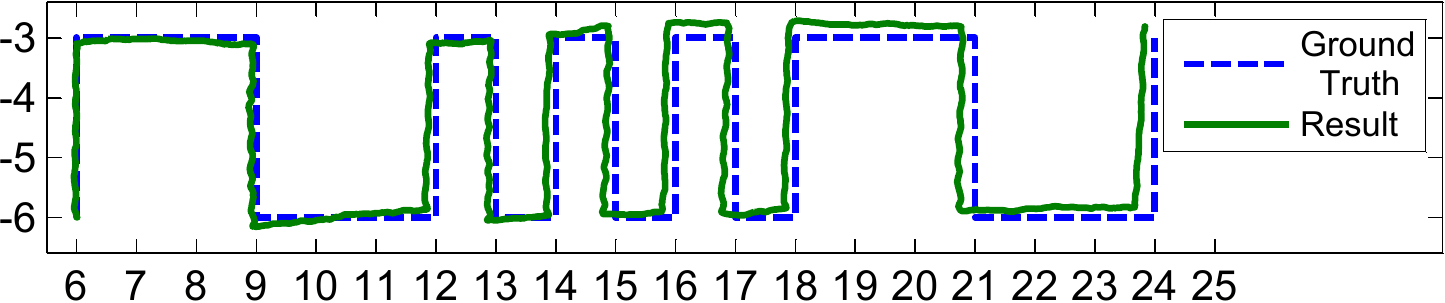}
    \end{center}
    \caption{Precise real-time indoor tracking.}
    \label{fig:realtime}
\end{figure}

We also conduct  real time indoor tracking using the same environment
as in Figure \ref{fig:indoorplacement}.
Assume that we  get the initial position of the user before s/he walks
by  shaking the phone.
In our experiments reported here, users starts from spot
$(6,-6)$ shown in Figure \ref{fig:realtime}. Then, the user
walks in some specific paths with length more than $50m$ with the
phone in his/her hand to the destination at spot $(24,-3)$. The errors are kept within $0.4m$ shown in Figure \ref{fig:trackingerrordetail} and Figure \ref{fig:realtime}.

We then consider the case that there are errors on the calculated initial
 position when the user starts walking.
For each test, we uniformly choose a spot which is $0.25m$,
 $0.5m$, $0.75m$, or $1m$ from $(6,-6)$, and
 measure the localization accuracies at the destination,
 \ie, distances from $(24,-6)$ to the calculated
 final positions in Figure \ref{fig:initialerroreffect}. We can observe that the errors at initial position do not affect the real
 time tracking, where the error is still within $2m$ when the user
 walks for 51 meters and the initial position error is $1m$.

As the phone needs $3.9s$ to process the acoustic samples of 1s, for
 real-time tracking by \ourprotocol,
 we let the phone process $20\%$ of the samples, instead
 of  full samples.
Specifically, it processes consecutive samples of
 $0.05s$ for each $0.25s$.
Hence, the phone can deal with the samples and
 track the position in real time.
The result is  close to the one which processes full  samples in Figure~\ref{fig:realtime}.
We  plot the localization errors in
Figure~\ref{fig:trackingerrordetail}.
The mean error and standard deviation in  this case is $0.29m$ and $0.34m$
respectively, which is still very precise. The CPU usage can also be
lowered down by using $10\%$ of the sample with the mean error of
$1.02m$, if the CPU of some other phone is not fast enough.

\section{Conclusion}
\label{sec:conclusion}

In this paper, we propose \ourprotocol, a novel acoustic-based method
 to find the direction of the acoustic source, and a real-time accurate
 indoor localization scheme based on this precise direction-finding.  
\ourprotocol effectively    leverages the Doppler effects of
 the acoustic waves received by phones by exploiting the 
 sensors in the smartphone and existing speakers to send sinusoidal
 signals. 
Our extensive evaluations show that \ourprotocol performs extremely
 well in phone-to-phone direction finding and real-time indoor
 localization.  Note that \ourprotocol did not directly use the ranging result as accurate ranging often needs either time-synchronization or communication between two nodes, both of which incur overhead. Some future work are to study the optimal placement of acoustic anchors,
 and to develop a low overhead distance estimation between phone and source for
 further improving the performances and reducing the number of anchors of \ourprotocol.

{\small
\bibliographystyle{acm}

\bibliography{RectangleDirection}

\begin{thebibliography}{10}

\bibitem{byteLight}
Bytelight technology.
\newblock \url{http://www.bytelight.com/}.

\bibitem{friendshake}
Facebook's friendshake.
\newblock \url{http://www.facebook.com}.

\bibitem{zaxis}
Getting the force of gravity by using the accelerometer.
\newblock
  \url{https://developer.android.com/reference/android/hardware/SensorEvent.html}.

\bibitem{GoogleLatitude}
Google latitude.
\newblock \url{https://www.google.com.hk/latitude/}.

\bibitem{DBLP:conf/infocom/BahlP00}
{\sc Bahl, P., and Padmanabhan, V.~N.}
\newblock Radar: An in-building rf-based user location and tracking system.
\newblock In {\em INFOCOM\/} (2000).

\bibitem{citeulike:7912602}
{\sc Beauregard, S., Haas, and Wpnc}.
\newblock {Pedestrian dead reckoning: A basis for personal positioning}.
\newblock {\em WPNC\/} (2006).

\bibitem{5168931}
{\sc Chandrasekaran, G., Ergin, M., Yang, J., Liu, S., Chen, Y., Gruteser, M.,
  and Martin, R.}
\newblock Empirical evaluation of the limits on localization using signal
  strength.
\newblock In {\em SECON\/} (2009).

\bibitem{claerbout1992earth}
{\sc Claerbout, J.}
\newblock {\em Earth soundings analysis: Processing versus inversion}.
\newblock Blackwell Scientific Publications, 1992.

\bibitem{2010-MOBICOM-Didyousee}
{\sc Constandache, I., Bao, X., Azizyan, M., and Choudhury, R.~R.}
\newblock Did you see bob?: human localization using mobile phones.
\newblock In {\em MobiCom\/} (2010).

\bibitem{citeulike:5657344}
{\sc Johnson, C.~R., and Sethares, W.~A.}
\newblock {\em Telecommunication Breakdown; Concepts of communication
  Transmitted via {Software-Defined} Radio}.
\newblock Prentice Hall, August 2003.

\bibitem{4711074}
{\sc Kim, M., and Chong, N.~Y.}
\newblock Direction sensing rfid reader for mobile robot navigation.
\newblock {\em Automation Science and Engineering, IEEE Transactions on\/}
  (2009).

\bibitem{DBLP:journals/cee/KulakowskiVELG10}
{\sc Kulakowski, P., Vales-Alonso, J., Egea-L{\'o}pez, E., Ludwin, W., and
  Garc\'{\i}a-Haro, J.}
\newblock Angle-of-arrival localization based on antenna arrays for wireless
  sensor networks.
\newblock {\em Computers {\&} Electrical Engineering\/} (2010).

\bibitem{2007-SenSys-Trackingmobilenodes}
{\sc Kusy, B., L{\'e}deczi, {\'A}., and Koutsoukos, X.~D.}
\newblock Tracking mobile nodes using rf doppler shifts.
\newblock In {\em SenSys\/} (2007).

\bibitem{2008-SenSys-Spinningbeaconsprecise}
{\sc lin Chang, H., ben Tian, J., Lai, T.-T., Chu, H.-H., and Huang, P.}
\newblock Spinning beacons for precise indoor localization.
\newblock In {\em SenSys\/} (2008).

\bibitem{DBLP:journals/tsmc/LiuDBL07}
{\sc Liu, H., Darabi, H., Banerjee, P.~P., and Liu, J.}
\newblock Survey of wireless indoor positioning techniques and systems.
\newblock {\em IEEE Transactions on Systems, Man, and Cybernetics, Part C\/}
  (2007).

\bibitem{2012-MOBICOM-PushlimitWiFi}
{\sc Liu, H., Gan, Y., Yang, J., Sidhom, S., Wang, Y., Chen, Y., and Ye, F.}
\newblock Push the limit of wifi based localization for smartphones.
\newblock In {\em MobiCom\/} (2012).

\bibitem{2012-MOBICOM-Centaurlocatingdevices}
{\sc Nandakumar, R., Chintalapudi, K.~K., and Padmanabhan, V.~N.}
\newblock Centaur: locating devices in an office environment.
\newblock In {\em MobiCom\/} (2012).

\bibitem{2003-INFOCOM-AdHocPositioning}
{\sc Niculescu, D., and Badrinath, B.~R.}
\newblock Ad hoc positioning system (aps) using aoa.
\newblock In {\em INFOCOM\/} (2003).

\bibitem{Niculescu:2004:VBS:1023720.1023727}
{\sc Niculescu, D., and Nath, B.}
\newblock Vor base stations for indoor 802.11 positioning.
\newblock In {\em MobiCom\/} (2004).

\bibitem{2012-MobiQuitous2011-ProposalDirectionEstimation}
{\sc Nishimura, Y., Imai, N., and Yoshihara, K.}
\newblock A proposal on direction estimation between devices using acoustic
  waves.
\newblock In {\em MobiQuitous\/} (2012).

\bibitem{2007-SenSys-BeepBeephighaccuracy}
{\sc Peng, C., Shen, G., Zhang, Y., Li, Y., and Tan, K.}
\newblock Beepbeep: a high accuracy acoustic ranging system using cots mobile
  devices.
\newblock In {\em SenSys\/} (2007).

\bibitem{2009-MobiSys-Point&Connectintentionbased}
{\sc Peng, C., Shen, G., Zhang, Y., and Lu, S.}
\newblock Point{\&}connect: intention-based device pairing for mobile phone
  users.
\newblock In {\em MobiSys\/} (2009).

\bibitem{Priyantha:2000:CLS:345910.345917}
{\sc Priyantha, N.~B., Chakraborty, A., and Balakrishnan, H.}
\newblock The cricket location-support system.
\newblock In {\em MobiCom\/} (2000).

\bibitem{6071927}
{\sc Prorok, A., Tome, P., and Martinoli, A.}
\newblock Accommodation of nlos for ultra-wideband tdoa localization in single-
  and multi-robot systems.
\newblock In {\em IPIN\/} (2011).

\bibitem{2011-SenSys-feasibilityrealtime}
{\sc Qiu, J., Chu, D., Meng, X., and Moscibroda, T.}
\newblock On the feasibility of real-time phone-to-phone 3d localization.
\newblock In {\em SenSys\/} (2011).

\bibitem{2012-MOBICOM-Zeezeroeffort}
{\sc Rai, A., Chintalapudi, K.~K., Padmanabhan, V.~N., and Sen, R.}
\newblock Zee: zero-effort crowdsourcing for indoor localization.
\newblock In {\em MobiCom\/} (2012).

\bibitem{rice2008digital}
{\sc Rice, M.}
\newblock {\em Digital Communications: A Discrete-Time Approach}.
\newblock Prentice Hall, 2008.

\bibitem{rosen2009encyclopedia}
{\sc Rosen, J., and Gothard, L.}
\newblock {\em Encyclopedia of Physical Science}.
\newblock 2009.

\bibitem{DBLP:journals/trob/SeLL05}
{\sc Se, S., Lowe, D.~G., and Little, J.~J.}
\newblock Vision-based global localization and mapping for mobile robots.
\newblock {\em IEEE Transactions on Robotics\/} (2005).

\bibitem{4509717}
{\sc Subramanian, A., Deshpande, P., Gaojgao, J., and Das, S.}
\newblock Drive-by localization of roadside wifi networks.
\newblock In {\em INFOCOM\/} (2008).

\bibitem{Wang:2012:NNW:2307636.2307655}
{\sc Wang, H., Sen, S., Elgohary, A., Farid, M., Youssef, M., and Choudhury,
  R.~R.}
\newblock No need to war-drive: unsupervised indoor localization.
\newblock In {\em MobiSys\/} (2012).

\bibitem{Want:1992:ABL:128756.128759}
{\sc Want, R., Hopper, A., Falc\~{a}o, V., and Gibbons, J.}
\newblock The active badge location system.
\newblock {\em ACM Trans. Inf. Syst.\/} (1992).

\bibitem{626982}
{\sc Ward, A., Jones, A., and Hopper, A.}
\newblock A new location technique for the active office.
\newblock {\em Personal Communications, IEEE\/} (1997).

\bibitem{DBLP:conf/mobicom/YangSCVLCCGM11}
{\sc Yang, J., Sidhom, S., Chandrasekaran, G., Vu, T., Liu, H., Cecan, N.,
  Chen, Y., Gruteser, M., and Martin, R.~P.}
\newblock Detecting driver phone use leveraging car speakers.
\newblock In {\em MobiCom\/} (2011).

\bibitem{2012-MOBICOM-Locatinginfingerprint}
{\sc Yang, Z., Wu, C., and Liu, Y.}
\newblock Locating in fingerprint space: wireless indoor localization with
  little human intervention.
\newblock In {\em MobiCom\/} (2012).

\bibitem{Youssef:2005:HWL:1067170.1067193}
{\sc Youssef, M., and Agrawala, A.}
\newblock The horus wlan location determination system.
\newblock In {\em MobiSys\/} (2005).

\bibitem{2012-MobiSys-SwordFightenablingnew}
{\sc Zhang, Z., Chu, D., Chen, X., and Moscibroda, T.}
\newblock Swordfight: enabling a new class of phone-to-phone action games on
  commodity phones.
\newblock In {\em MobiSys\/} (2012).

\bibitem{2011-MOBICOM-Iamantenna}
{\sc Zhang, Z., Zhou, X., Zhang, W., Zhang, Y., Wang, G., Zhao, B.~Y., and
  Zheng, H.}
\newblock I am the antenna: accurate outdoor ap location using smartphones.
\newblock In {\em MobiCom\/} (2011).

\end{thebibliography}
}

\end{document}